\documentclass[aps,twocolumn,prl,superscriptaddress,footinbib]{revtex4-1}
\usepackage{amsmath,amssymb,bm}
\usepackage{graphicx}
\usepackage{epstopdf}
\usepackage{latexsym}
\usepackage[normalem]{ulem} 
\usepackage[usenames,dvipsnames]{color}
\usepackage{hyperref}
\usepackage{natbib}
\usepackage{braket}
\begin{document}

\newcommand{\Rev }[1]{{\color{black}{#1}\normalcolor}} % Revision
\newcommand{\RevSec }[1]{{\color{black}{#1}\normalcolor}} % Revision
\newcommand{\Com}[1]{{\color{red}{#1}\normalcolor}} %Comment
\newcommand{\Junk}[1]{{\color{magenta}{#1}\normalcolor}} %Filler text

\title{\Rev{Detecting out-of-time-order correlations via quasi-adiabatic echoes as a tool to reveal quantum coherence in equilibrium quantum phase transitions}}
\date{\today}

\author{R.~J. Lewis-Swan}
\affiliation{Homer L. Dodge Department of Physics and Astronomy, The University of Oklahoma, Norman, Oklahoma 73019, USA}
\affiliation{JILA, NIST, Department of Physics, University of Colorado, Boulder, CO 80309, USA}
\affiliation{Center for Theory of Quantum Matter, University of Colorado, Boulder, CO 80309, USA}
\author{S.~R. Muleady}
\affiliation{JILA, NIST, Department of Physics, University of Colorado, Boulder, CO 80309, USA}
\affiliation{Center for Theory of Quantum Matter, University of Colorado, Boulder, CO 80309, USA}
\author{A.~M. Rey}
\affiliation{JILA, NIST, Department of Physics, University of Colorado, Boulder, CO 80309, USA}
\affiliation{Center for Theory of Quantum Matter, University of Colorado, Boulder, CO 80309, USA}

\begin{abstract}
\Rev{We propose a new dynamical method to connect equilibrium quantum phase transitions and quantum coherence using out-of-time-order correlations (OTOCs). Adopting the iconic Lipkin-Meshkov-Glick and transverse-field Ising models as illustrative examples, we show that an abrupt change in coherence and entanglement of the ground state across a quantum phase transition is observable in the spectrum of multiple quantum coherence (MQC) intensities, which are a special type of OTOC. We also develop a robust protocol to obtain the relevant OTOCs using quasi-adiabatic quenches through the ground state phase diagram. Our scheme allows for the detection of OTOCs without time-reversal of coherent dynamics, making it applicable and important for a broad range of current experiments where time-reversal cannot be achieved by inverting the sign of the underlying Hamiltonian.}
\end{abstract}

\maketitle

\noindent {\it Introduction:} Quantum phase transitions (QPTs) \cite{sachdev_QPT_2007} play a central role in many fields of quantum science and have been studied using a variety of tools. Fundamentally, a QPT is signaled \Rev{in the thermodynamic limit} by a non-analyticity in the energy density of the ground state, or a vanishing energy gap between the ground and lowest-excited states. However, quantum information science has ignited a theoretical push towards characterizing the critical region of a QPT through information-theoretic quantities such as entanglement entropy \cite{Osborne_2002,Vidal_2003,Eisert_2010}, Loschmidt echoes \cite{Quan_2006,Zanardi_2006}, fidelity susceptibility \cite{Gu_2010,Capponi_2010,Troyer_2015}, and coherence measures \cite{Chen_2016,Karpat_2014,Li_2016_coherence,Li_2020_coherence,Hu_2020}.

\Rev{
In parallel, the past two decades have seen a growing focus on the dynamics of quantum information \cite{RLS_2019quantuminfo} and non-equilibrium systems \cite{polkovnikov_2011review} as a result of improvements in the technical capabilities of atomic, molecular, and optical (AMO) experiments. Of recent note is the study of quantum chaos and information scrambling using out-of-time-order correlations (OTOCs) \cite{Hayden_2007,Sekino2008,Shenker2014,Kitaev2015}. Already, OTOCs have been adapted to study phenomena beyond their original purview, including the diagnosis of dynamical \cite{Heyl_2018,Xu_QPT_OTOC_2020} and equilibrium phase transitions \cite{Shen_2017,Duan_2019,Li_QPT_OTOC_2020}. \Rev{Recent work has also demonstrated that a special type of fidelity OTOC (FOTOC) \cite{RLS_2019scrambling} can be a powerful tool to diagnose entanglement \cite{Martin_2018} and coherence \cite{Martin2017_OTOC} in many-body systems through the framework of multiple quantum coherences (MQCs) pioneered in NMR \cite{Baum_1985,Munowitz_1987,cappellaro_2014}. 

Motivated by these developments, here we demonstrate that FOTOCs and the spectrum of MQC intensities can serve as a unifying tool to connect concepts of quantum coherence to equilibrium QPTs in a dynamical setting. In doing so, we also develop a powerful new protocol to dynamically access FOTOCs, distinct from recent schemes in that we do not require time-reversal via altering the sign of the Hamiltonian \cite{RLS_2019scrambling,Martin2017_OTOC,Li_2017,Gadway_2017,PhysRevLett.120.070501,PhysRevA.94.062329}, nor do we demand auxiliary qubits \cite{Swingle_2016,1607.01801,Landsman2019} or exhaustive measurements \cite{Molmer_2020,PhysRevX.9.021061,2001.02176}.}

In this manuscript, we utilize the paradigmatic Lipkin-Meshkov-Glick (LMG) and transverse-field Ising (TFI) models as case studies to demonstrate that FOTOCs can diagnose the non-analyticity of a QPT and can distinguish quantum phases even in the limit of small system size. We illustrate the power of FOTOCs as a practical tool for the characterization of QPTs by proposing a completely general, experimentally realistic dynamical protocol to obtain these correlations and the related MQC spectrum from a pseudo-echo of quasi-adiabatic dynamics. The technical simplicity of our scheme relative to the aforementioned alternatives means that our results are relevant for a broad range of experimentally accessible models in AMO and condensed matter physics in addition to the spin models studied here, with immediate impact for state-of-the-art quantum simulators. For example, relaxation of the time-reversal constraint opens the possibility to more easily study OTOCs in short-range Ising models simulated with trapped-ions, similar to the TFI model we investigate.
}

\noindent {\it Quantifying quantum coherence:} \Rev{The FOTOCs we study are defined as $F_{\phi} = \mathrm{Tr}\left[ \hat{W}^{\dagger}_{\phi}(t) \hat{\rho}_0 \hat{W}_{\phi}(t) \hat{\rho}_0 \right]$ where $\hat{W}(t) = \hat{U}^{\dagger}(t)e^{-i\phi\hat{A}}\hat{U}(t)$, $\hat{A}$ is a Hermitian operator, and $\hat{U}(t)$ describes a unitary time-evolution operator that will be specified later in the manuscript. We connect FOTOCs to MQCs \cite{Martin_2018,RLS_2019scrambling} using cyclicity of the trace to rewrite $F_{\phi} = \mathrm{Tr}\left[ \hat{\rho} \hat{\rho}^{\phi} \right]$, where we define $\hat{\rho} = \hat{U}(t)\hat{\rho}_0\hat{U}^{\dagger}(t)$ and $\hat{\rho}^{\phi} \equiv e^{-i\phi\hat{A}} \hat{\rho} e^{i\phi\hat{A}}$. We have dropped the explicit time dependence on $\hat{\rho}$ to simplify notation. The Fourier transform of the FOTOC $F_{\phi}$ then defines the spectrum of MQC intensities: $I^{\hat{A}}_m(\hat{\rho}) = \sum_{\phi} F_{\phi} e^{im\phi}$. 

The MQC intensities are a well-established signature of the coherence of a many-body state \cite{Baum_1985,Munowitz_1987,cappellaro_2014,Garrahan_2019}. This can be seen using the alternative definition \cite{Martin_2018} $I^{\hat{A}}_m(\hat{\rho}) \equiv \mathrm{Tr}\left[ \hat{\rho}_{-m} \hat{\rho}_m \right]$ where $\hat{\rho}_m = \sum_{\lambda_i - \lambda_j = m} \rho_{ij} \vert \lambda_i \rangle \langle \lambda_j \vert$ with $\vert \lambda_i \rangle$ the eigenstates of a given Hermitian operator $\hat{A}$ such that $\hat{A}\vert \lambda_i \rangle = \lambda_i \vert \lambda_i \rangle$. By construction, the blocks $\hat{\rho}_m$ contain all coherences between eigenstates of $\hat{A}$ differing by $m$, which is then quantified via $I^{\hat{A}}_m(\hat{\rho})$.}

\noindent {\it \Rev{Signals of a QPT in MQCs:}} \Rev{Here we show that the spectrum of MQC intensities, accessible via FOTOCs, allows a robust and experimentally accessible characterization of QPTs in terms of quantum coherence \cite{Chen_2016}. Our central insight is that the drastic change in the coherence and entanglement of a many-body ground state across a QPT should lead to a correspondingly sharp change in the features of the MQC spectrum for an appropriately chosen $\hat{A}$.}

To formalize this statement, we consider a \Rev{general} toy Hamiltonian: $\hat{H} = \hat{H}_1 + \lambda\hat{H}_2$, where $\left[\hat{H}_1,\hat{H}_2\right] \neq 0$ and $\lambda \geq 0$ is a dimensionless (tunable) parameter. We take $\hat{A} = \hat{H}_2$ and consider the ground state in the limiting cases $\lambda \to 0$ and $\lambda \to \infty$. In the latter case, the ground state $\vert \psi^{\lambda\to\infty}_{\mathrm{GS}}\rangle$ of $\hat{H}$ is the lowest-energy eigenstate of $\hat{H}_2$, and thus $\hat{\rho}^{\lambda\to\infty}_{\mathrm{GS}} = \vert \psi^{\lambda\to\infty}_{\mathrm{GS}}\rangle \langle \psi^{\lambda\to\infty}_{\mathrm{GS}} \vert$ is composed of a single diagonal entry in the eigenbasis defined by $\hat{A} = \hat{H}_2$. Trivially, the MQC spectrum will then be composed of a single peak, $I^{\hat{H}_2}_m(\hat{\rho}^{\lambda\to\infty}_{\mathrm{GS}}) = \delta_{m,0}$, due to the lack of coherences with respect to $\hat{H}_2$. Conversely, for $\lambda \to 0$ the ground state $\vert \psi^{\lambda\to0}_{\mathrm{GS}}\rangle$ becomes an eigenstate of $\hat{H}_1$. As $\left[\hat{H}_1,\hat{H}_2\right] \neq 0$, this ground state cannot be a simultaneous eigenstate of $\hat{H}_2$: $\vert \psi^{\lambda\to0}_{\mathrm{GS}}\rangle$ must be a coherent superposition of eigenstates of $\hat{H}_2$ such that the density matrix $\hat{\rho}^{\lambda\to0}_{\mathrm{GS}} = \vert \psi^{\lambda\to0}_{\mathrm{GS}}\rangle \langle \psi^{\lambda\to0}_{\mathrm{GS}} \vert$ possesses off-diagonal coherences with respect to the $\hat{H}_2$ eigenbasis. Consequently, we expect a (relatively) broad MQC spectrum with non-zero $I^{\hat{H}_2}_{m}(\hat{\rho}^{\lambda\to0}_{\mathrm{GS}})$ for $m\neq 0$.

\Rev{Extending the spirit of this argument between these two limits, we expect for a model possessing a QPT at a critical point $\lambda_c$ that the transition will generically be signaled by an abrupt change from a narrow to broad MQC spectrum. \RevSec{This expectation can also be supported by the fact that the spectral width $2\sigma^2_{\mathrm{MQC}} = 2\sum_m m^2 I^{\hat{H}_2}_m(\hat{\rho}_{\mathrm{GS}})$ is a lower bound for the quantum Fisher information (QFI) \cite{Braunstein1994,Martin_2018,Garrahan_2019} of the state $\rho_{\mathrm{gs}}$ with respect to $\hat{H}_2$ (saturated for pure states)}. Prior work has shown the QFI can provide signatures of a QPT \cite{Hauke2016} although this was typically computed with respect to a known order parameter rather than $\hat{H}_2$. We proceed to show that not only the width of this MQC spectrum but also, more importantly, individual intensities themselves can signal a QPT. An experimentally relevant quantity is $I^{\hat{H}_2}_0$, which features a divergent derivative $\frac{d^2}{d\lambda^2}I^{\hat{H}_2}_0$ at the QPT in the thermodynamic limit.}

\begin{figure}[!]
 \includegraphics[width=8cm]{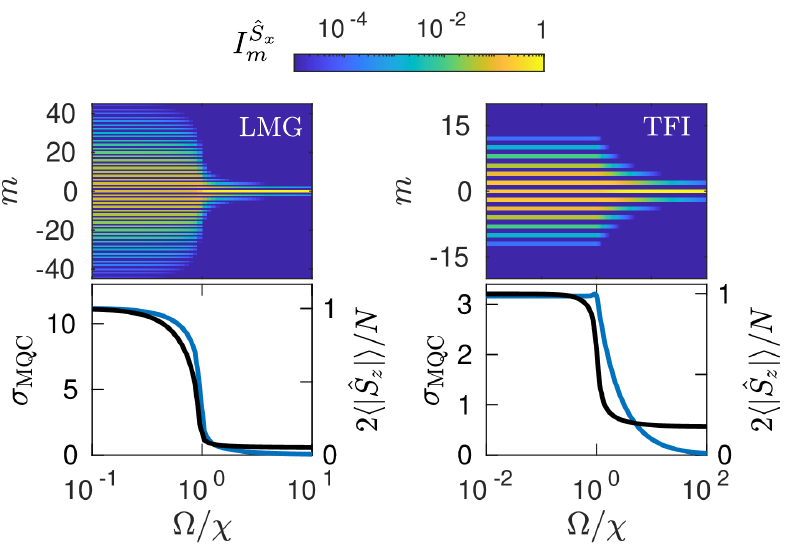}
 \caption{Characteristic MQC spectra $I^{\hat{S}_x}_m(\hat{\rho}_{\mathrm{GS}})$ of the numerically-computed ground state for the LMG ($N=250$) and the analytically-computed ground state for the TFI ($N=20$) models as a function of $\Omega/\chi$. The phase boundary near $\Omega/\chi \approx 1$ in both models is signified by an abrupt change in the spectrum width $\sigma_{\mathrm{MQC}}$ (blue line, lower panels), which we also compare to the order parameter $\langle \vert\hat{S}_z\vert \rangle$ (black, lower panels).}
 \label{fig:MQCdist}
\end{figure}

\noindent {\it \Rev{Demonstrative examples:}} We illustrate and validate our arguments with a pair of iconic models of quantum magnetism: the anisotropic LMG \cite{LMG_1965,Vidal_LMG_2007} and the one-dimensional TFI \cite{DeGennes_TFI_1963,Stinchcombe_1973,TFIbook_2015} models. The former describes an ensemble of $N$ mutually interacting spin-$1/2$s in a transverse field, while the latter involves only nearest-neighbour interactions. Each model can be described within the general Hamiltonian
\begin{equation}
 \hat{H} = -\frac{1}{2\mathcal{C}} \sum_{i < j} \chi_{ij}\hat{\sigma}^z_i\hat{\sigma}^z_{j} - \frac{\Omega}{2}\sum_i \hat{\sigma}^x_i , \label{eqn:H} 
\end{equation}
where $\hat{\sigma}^{\alpha}_i$ are Pauli operators for site $i$ and $\alpha = x,y,z$. The interaction between spins at sites $i$ and $j$ is characterized by $\chi_{ij}$, and $\Omega$ is the transverse field strength. We normalize the interaction by $\mathcal{C} = \left(\sum_{i,j} \chi_{ij}\right)/\chi N$; for the LMG model $\chi_{ij} = \chi$, while for the TFI model $\chi_{ij} = \chi\delta_{i,j-1}$. We adopt $\hbar = 1$ throughout the manuscript.

Each model features a second-order QPT between ferromagnetic and paramagnetic phases at a critical point $(\Omega/\chi)_c = 1$. The ground state physics can be described within the basis of fully symmetric spin states $\vert m_{\alpha} \rangle$ defined by $\hat{S}_{\alpha}\vert m_{\alpha} \rangle = m_{\alpha} \vert m_{\alpha}\rangle$, where $\hat{S}_{\alpha} \equiv \sum_j (\hat{\sigma}^{\alpha}_j/2)$ and we suppress the quantum number $S = N/2$ for brevity. In the strong-field limit $\Omega/\chi \gg 1$, the paramagnetic ground state is characterized by the polarization of all spins identically along $\hat{x}$, $\vert \psi^{P}_{\mathrm{GS}} \rangle = \vert (N/2)_x \rangle$, while in the weak-field limit $\Omega/\chi \ll 1$, the ferromagnetic ground state is an entangled GHZ state $\vert \psi^{F}_{\mathrm{GS}} \rangle = \vert (N/2)_z \rangle \pm \vert -(N/2)_z \rangle$. We work with the symmetric state, which is adiabatically connected to the paramagnetic ground state for finite $N$.

\begin{figure}[tb!]
 \includegraphics[width=8cm]{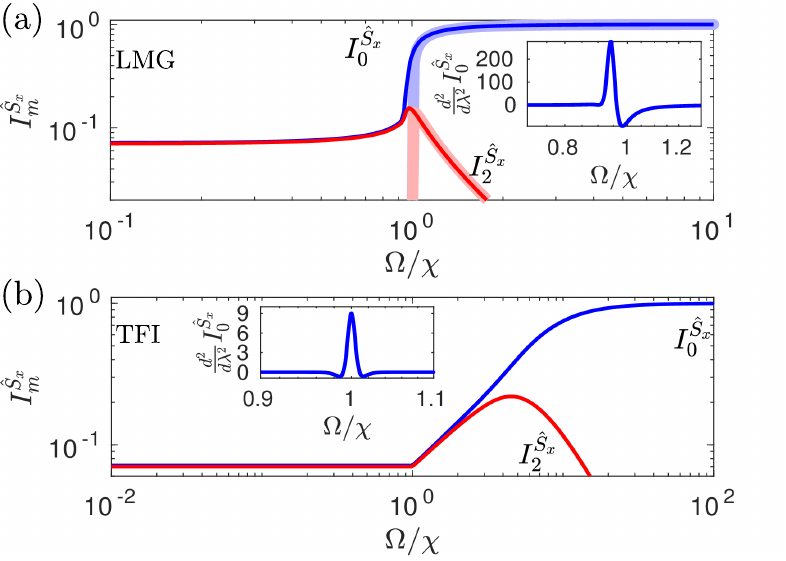}
 \caption{Signatures of a QPT in the MQC intensities $I^{\hat{S}_x}_m$. (a) The LMG model in the $N\to\infty$ limit (faded lines) has abruptly vanishing intensities $I^{\hat{S}_x}_0$ (blue) and $I^{\hat{S}_x}_2$ (red) at the critical point, $(\Omega/\chi)_c = 1$. Dark lines indicate a numerical comparison for $N=250$. (b) The TFI model displays a sharp kink in the $I^{\hat{S}_x}_0$ and $I^{\hat{S}_x}_2$ components at the critical point $(\Omega/\chi)_c = 1$. We plot results for the analytically-computed ground state using \Rev{$N=250$ to facilitate comparison with the LMG results. Insets for (a)-(b) highlight the divergence in $d^2I^{\hat{S}_x}_0/d\Omega^2$ at the QPT for $N=250$.}}
 \label{fig:Im}
\end{figure}

We demonstrate that the MQC spectrum can serve as a diagnostic tool to distinguish equilibrium phases by computing the intensities with respect to the transverse field term in Eq.~(\ref{eqn:H}), $\hat{A} = \hat{S}_x$. In Fig.~\ref{fig:MQCdist} we plot the distribution as a function of $\Omega/\chi$ for $N=250$ (LMG) and $N=20$ (TFI). In both cases, deep in the paramagnetic phase the spectrum is dominated by a sharp peak at $I_0 \simeq 1$, reflecting that the ground state $\vert \psi^{P}_{\mathrm{GS}} \rangle$ lacks coherences with respect to the eigenstates of $\hat{S}_x$. Conversely, coherences are generated as the transverse field is reduced relative to the interactions, with the QPT of each model reflected in the abrupt growth of the width $\sigma_{\mathrm{MQC}}$ of the MQC spectrum near $(\Omega/\chi)_c=1$. For the TFI model the change in $\sigma_{\mathrm{MQC}}$, approaching from $\Omega/\chi < 1$, is sharp even for this small system ($N=20$ in Fig.~1), and is a clearer indication of the transition than the associated order parameter $\langle \vert \hat{S}_z \vert \rangle$. 

\Rev{Besides the growth of the width, we are also able to establish both analytically and numerically that the QPT is signaled directly in the individual MQC intensities (see Ref.~\cite{SM} for relevant expressions). Our most relevant observation is that the derivative $d^2I^{\hat{S}_x}_0(\hat{\rho}_{\mathrm{GS}})/d\Omega^2$ diverges at $\Omega_c$ as shown in Fig.~2. A similar feature is observable in $d^2I^{\hat{S}_x}_2(\hat{\rho}_{\mathrm{GS}})/d\Omega^2$ (\cite{SM}).} 

\Rev{For the LMG model, the location of the transition $(\Omega/\chi)_*$, taken as the peak of $\frac{d^2}{d\Omega^2}I^{\hat{S}_x}_0(\hat{\rho}_{\mathrm{GS}})$ for a finite system, approaches the $N\to\infty$ critical point as $[1 - (\Omega/\chi)_*] \sim N^{-0.65}$ \cite{SM}, consistent with that of $(\Omega/\chi)_*$ obtained from the peak in the susceptibility $\frac{d}{d\Omega}\langle \vert \hat{S}_z \vert \rangle$ over the same window of system size $N$. Similarly, for the TFI model the location of the peak in $\frac{d^2}{d\Omega^2}I_0^{\hat{S}_x}(\hat{\rho}_{\mathrm{GS}})$ approaches as $[1 - (\Omega/\chi)_*] \sim N^{-2}$ \cite{SM,Damski_2013}. This verifies that the MQC signatures do not display any systematic offset from the QPT beyond finite size effects \cite{ZurekBoseHubbard}.}

\Rev{We also point out that the LMG model predicts a sharp peak in $I^{\hat{S}_x}_{m>0}(\hat{\rho}_{\mathrm{GS}})$ in the paramagnetic phase close to the QPT. This could serve as a more modest proxy for the QPT in experimental systems \cite{SM}.} Conversely, the TFI model shows a similar peak but it shifts further into the paramagnetic phase as $N$ increases and is uncorrelated with the QPT. 

The demonstration that the QPT is unambiguously signaled by an abrupt change of the spectral width and character of individual MQC intensities in both of these models, regardless of the fact that they belong to different universality classes, emphasizes the general utility of the MQC spectrum to diagnose a QPT.

\begin{figure*}[tb!]
 \includegraphics[width=16cm]{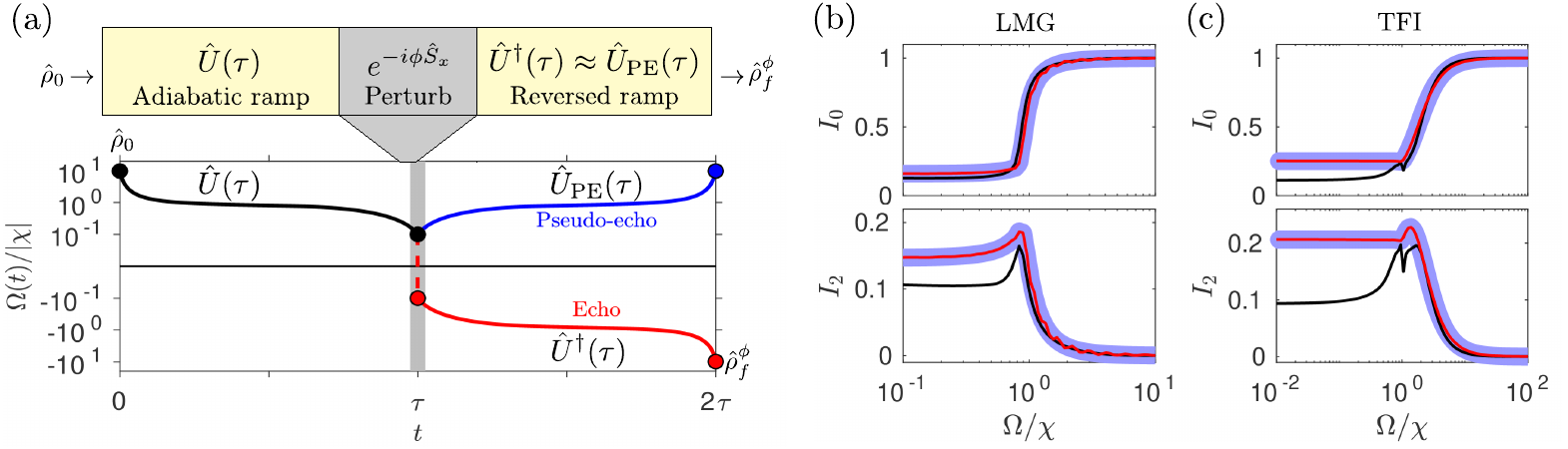}
 \caption{(a) Schematic of many-body echo to obtain the MQC spectrum. The system is initialized in $\hat{\rho}_0$, corresponding to the ground state at $\Omega(0)/\chi$, before the field is slowly ramped to $\Omega(\tau)$, described by unitary $\hat{U}(\tau)$. A global rotation is imprinted on the state before the dynamics are reversed via i) a many-body echo $\hat{U}^{\dagger}(t)$ or ii) a pseudo-echo $\hat{U}_{\mathrm{PE}}(\tau)$. In the former the sign of the Hamiltonian is also flipped, whereas in the latter it is not. (b)-(c) Benchmark of dynamical protocol. The MQC components predicted from the exact ground state (faded blue lines) compared to those obtained from a pseudo-echo ramping sequence of duration $\chi\tau = 10$ (black) and $\chi\tau = 100$ (red). All data is from numerical simulations using $N=50$ and $N=20$ for the LMG and TFI models, respectively (see Ref.~\cite{SM} for details of ramps). 
 }
 \label{fig:DynamicalIm}
\end{figure*}

\Rev{
\noindent {\it Obtaining FOTOCs without time-reversal:} \Rev{The measurement of FOTOCs and the associated MQC spectrum has previously been achieved via echoes based on the time-reversal of coherent dynamics \cite{Martin2017_OTOC,Li_2017}. This can be seen from the definition of the FOTOC given previously, $F_{\phi} = \mathrm{Tr}\left[ \hat{W}^{\dagger}_{\phi}(t) \hat{\rho}_0 \hat{W}_{\phi}(t) \hat{\rho}_0 \right]$ where $\hat{W}_{\phi}(t) = \hat{U}^{\dagger}(t)e^{-i\phi\hat{A}}\hat{U}(t)$, which is interpreted as a dynamical sequence starting from the state $\hat{\rho}_0$ followed by: i) unitary evolution $\hat{U}(t)$, ii) a perturbation $e^{-i\phi\hat{A}}$, iii) time-reversed dynamics $\hat{U}^{\dagger}(t)$, and iv) a projective measurement of the final overlap with $\hat{\rho}_0$. Here, we discuss an approach to obtain the MQC spectrum of a generic ground state $\hat{\rho}_{\mathrm{GS}}$ by a similar echo sequence that replaces the time reversal step with adiabatic dynamics. This substitution opens the possibility of accessing OTOCs in a much broader range of physical platforms than previously considered.}

\Rev{The proposed echo sequence is shown in Fig.~\ref{fig:DynamicalIm}(a), where we concretely identify $\hat{U}(t)$ as describing a slow ramp of the Hamiltonian parameters in time. While we stress that our protocol is entirely generic, for simplicity we focus here on the LMG and TFI models, which we later investigate numerically.} Our sequence starts from the state $\hat{\rho}(0) = \hat{\rho}^P_{\mathrm{GS}}$ prepared at large $\Omega(0)/\chi \gg 1$ (for simplicity we assume $\chi$ is fixed). The ratio of these parameters is slowly changed such that the instantaneous state of the system follows the ground state of the instantaneous Hamiltonian $\hat{H}(t)$ characterized by $\Omega(t)/\chi$, $\hat{\rho}(t) = \hat{U}(t) \hat{\rho}_0 \hat{U}^{\dagger}(t) \equiv \hat{\rho}_{\mathrm{GS}}(t)$. \Rev{A global rotation, $e^{-i\phi\hat{S}_x}$, is then imprinted on the state by, e.g., suddenly quenching to $\Omega(t)/\chi \gg 1$. Ideally, the echo is completed by flipping the sign of the instantaneous Hamiltonian such that $\hat{H}(t) \to -\hat{H}(2\tau - t)$ for $t>\tau$, where $\tau$ is the duration of the initial ramping sequence. However, the ability to control the sign of the Hamiltonian is out of reach in many state-of-the-art experiments.}

\Rev{To overcome this barrier and make our protocol more widely applicable, we propose that the MQC spectrum can still be exactly obtained using a \emph{pseudo-echo} ($\hat{U}_{\mathrm{PE}}(t)$ in Fig.~\ref{fig:DynamicalIm}(a)) where the sign of the Hamiltonian is \emph{not} flipped, as long as the ramping sequence is sufficiently adiabatic. Finally, the overlap with $\hat{\rho}^P_{\mathrm{GS}}$ is obtained to yield $F_{\phi}$. We justify the pseudo-echo by the observation that the state produced by sequential adiabatic ramps of the transverse field from $\Omega(0) \to \Omega(t) \to \Omega(0)$ is effectively identical to that produced by true time-reversal, \textit{i.e.} the system will return to the initial ground state up to an overall irrelevant phase (see Ref.~\cite{SM}).}

This is a significant result as it demonstrates that effective time-reversal can still be achieved in the absence of any control of the sign of the Hamiltonian. Moreover, we are able to extract detailed information about the coherences of the complex many-body state $\hat{\rho}(t)$ without technically challenging state tomography \cite{Haffner_2005,Gross_2010tomography,Lanyon_2017tomography,Torlai_2018} or randomized measurements \cite{Elben_2019,Brydges_2019}. Instead, the coherent dynamics in the second half of the echo protocol map this information to a relatively manageable measurement (of the overlap with a simple product state \cite{Bollinger_2018}).
}

\Rev{
\noindent {\it Numerical study of quasi-adiabatic ramps:} Realistically, technical constraints and decoherence preclude the possibility of truly adiabatic dynamics in an experiment. To address this issue, we investigate the robustness of the MQC spectrum to diabatic excitations generated in a realistic ramp of finite duration and demonstrate that it retains reliable signatures of the QPT} \footnote{As our consideration of non-adiabatic ramps is specifically motivated by limitations on ramp duration due to decoherence, we do not directly consider the effects of decoherence on the MQC spectrum in detail. A brief discussion of the expected relevance of decoherence may instead be found in \cite{SM}.}.

In Fig.~\ref{fig:DynamicalIm}(b)-(c) we present data for the previously discussed LMG and TFI models for ramps of duration \Rev{$\chi\tau = (10,100)$} starting in the paramagnetic phase. \Rev{In both cases we obtain the dynamics by numerically solving a time-dependent Schr\"{o}dinger equation for small systems.} For the LMG model we take $N=50$ and $\Omega(0) = 10$, while for the TFI model $N=20$ and $\Omega(0) = 10^2$, with both ramps tailored to reach $\Omega(\tau) = 10^{-2}$ \cite{SM}. \Rev{We focus on the low-$m$ individual intensities rather than the width $\sigma_{\mathrm{MQC}}$, as decoherence will typically make accessing large-$m$ intensities, and thus the full spectrum, technically challenging.} An important figure of merit is the fidelity with which the targeted ground state is prepared, $\mathcal{F} = \vert \langle \psi_{\mathrm{GS}}(\tau) \vert \psi(\tau) \rangle \vert^2$. For the ramp durations in Fig.~\ref{fig:DynamicalIm}(b)-(c), the fidelities are $\mathcal{F} = (0.13, 0.99)$ and $\mathcal{F} = (0.5, 0.99)$ for the LMG and TFI models, respectively. Even for low fidelity ramps, the intensities obtained from the pseudo-echo reasonably follow the predictions of the exact ground state \cite{SM}, despite the generation of appreciable low-energy excitations. As a consequence, when interpreting connections between the MQC spectrum and the QPT one should demonstrate that the physics observed is dominated by the $T=0$ ground state. One way to quantify this is to measure a return fidelity after a pseudo-echo in the absence of the perturbation, which is associated with $\mathcal{F}_{\phi=0}$ \cite{SM}. A large return fidelity approaching unity indicates the MQC spectrum is dominated by the ground state contribution.

\Rev{
\noindent {\it Experimental implementation:} Our proposal to diagnose equilibrium QPTs using quantum coherence can be implemented in a range of experimental platforms featuring sufficient control of quasi-adiabatic dynamics. Promising directions include trapped-ion quantum simulators of interacting spin models, and \Rev{quantum simulators of Hubbard-like models implemented with neutral/Rydberg atoms in optical lattices aided by a quantum gas microscope \cite{Gross_2017review} or in tweezer arrays \cite{Keesling_2019,Browaeys_2020}}. Trapped-ions \cite{Richerme2013_PRA,Jurcevic2017,Zhang2017,Martin2017_OTOC,Bollinger_2018,Wall2017} in particular can be used to simulate the TFI and LMG examples studied here in addition to more generic models with power-law spin-spin interactions mediated by phonons \cite{SM}. These systems feature sufficiently low decoherence rates as to enable both the high-quality preparation of the entangled ground state \cite{Bollinger_2018,Bollinger_2018_arxiv,SM} and the ability to measure state overlap \cite{Martin2017_OTOC}. While time-reversal has been demonstrated in Penning traps in the limit of all-to-all interactions, which are mediated by coupling to a single phonon mode \cite{Martin2017_OTOC}, the Hamiltonian sign cannot be controlled for more generic interactions mediated by multiple phonon modes and thus the pseudo-echo protocol we develop in this manuscript is of immediate relevance.
}

\noindent {\it Conclusion:} We have proposed and investigated a \Rev{dynamical} method to diagnose signatures of \Rev{quantum coherence in} a QPT \Rev{using FOTOCs}. Our approach is robust, \Rev{has modest technical demands}, and does not require time-reversal through control of the sign of the Hamiltonian. While our \Rev{numerical} investigation focused on spin models implementable in arrays of trapped-ions \Rev{and Rydberg atoms}, our results are broadly applicable to a range of AMO quantum simulators where ground state physics can be studied in a controlled and isolated environment.

\begin{acknowledgments}
\noindent{\textit{Acknowledgements:}} We acknowledge helpful discussions with Cindy Regal, Mark Brown, and Itamar Kimchi. This work is supported by the AFOSR grant FA9550-18-1-0319, by the DARPA and ARO grant W911NF-16-1-0576, the ARO single investigator award W911NF-19-1-0210, the NSF PHY1820885, NSF JILA-PFC PHY-1734006 grants, and by NIST. 
\end{acknowledgments}

\end{document}

\newcommand{\Rev }[1]{{\color{black}{#1}\normalcolor}} % Revision
\newcommand{\RevSec }[1]{{\color{black}{#1}\normalcolor}} % Revision
\newcommand{\Com}[1]{{\color{red}{#1}\normalcolor}} %Comment
\newcommand{\Junk}[1]{{\color{magenta}{#1}\normalcolor}} %Filler text

\renewcommand{\theequation}{S\arabic{equation}}
\renewcommand{\thefigure}{S\arabic{figure}}

\title{Supplemental Material: Detecting out-of-time-order correlations via quasi-adiabatic echoes as a tool to reveal quantum coherence in equilibrium quantum phase transitions}
\date{\today}

\author{Robert J. Lewis-Swan}
\affiliation{\RevSec{Homer L. Dodge Department of Physics and Astronomy, The University of Oklahoma, Norman, Oklahoma 73019, USA}}
\affiliation{JILA, NIST, Department of Physics, University of Colorado, Boulder, CO 80309, USA}
\affiliation{Center for Theory of Quantum Matter, University of Colorado, Boulder, CO 80309, USA}
\author{Sean R. Muleady}
\affiliation{JILA, NIST, Department of Physics, University of Colorado, Boulder, CO 80309, USA}
\affiliation{Center for Theory of Quantum Matter, University of Colorado, Boulder, CO 80309, USA}
\author{Ana Maria Rey}
\affiliation{JILA, NIST, Department of Physics, University of Colorado, Boulder, CO 80309, USA}
\affiliation{Center for Theory of Quantum Matter, University of Colorado, Boulder, CO 80309, USA}

\maketitle

\section{Quasi-adiabatic ramp scheme}
In Fig.~3 of the main text we present the multiple quantum coherence (MQC) spectrum of low-energy states obtained as a result of a quasi-adiabatic ramping sequence starting from the paramagnetic ground state $\hat{\rho}^P_{\mathrm{GS}}$ at large field $\Omega/\chi \gg 1$. To be more specific, we initialize in the fully polarized state $\hat{\rho}^P_{\mathrm{GS}} = \vert (N/2)_x \rangle \langle (N/2)_x \vert$ at an initial transverse field $\Omega(0)$ and fixed interaction strength $\chi = 1$. We slowly quench the transverse field over a duration $\tau$ down to $\Omega(\tau)$ according to a local adiabatic approximation (LAA) \cite{Richerme_2013,Balasubramanian_2018} to a final value $\Omega(\tau)$. The MQC spectrum is then obtained via a pseudo-echo which is detailed in the following section.

\begin{figure}[tb!]
 \includegraphics[width=8cm]{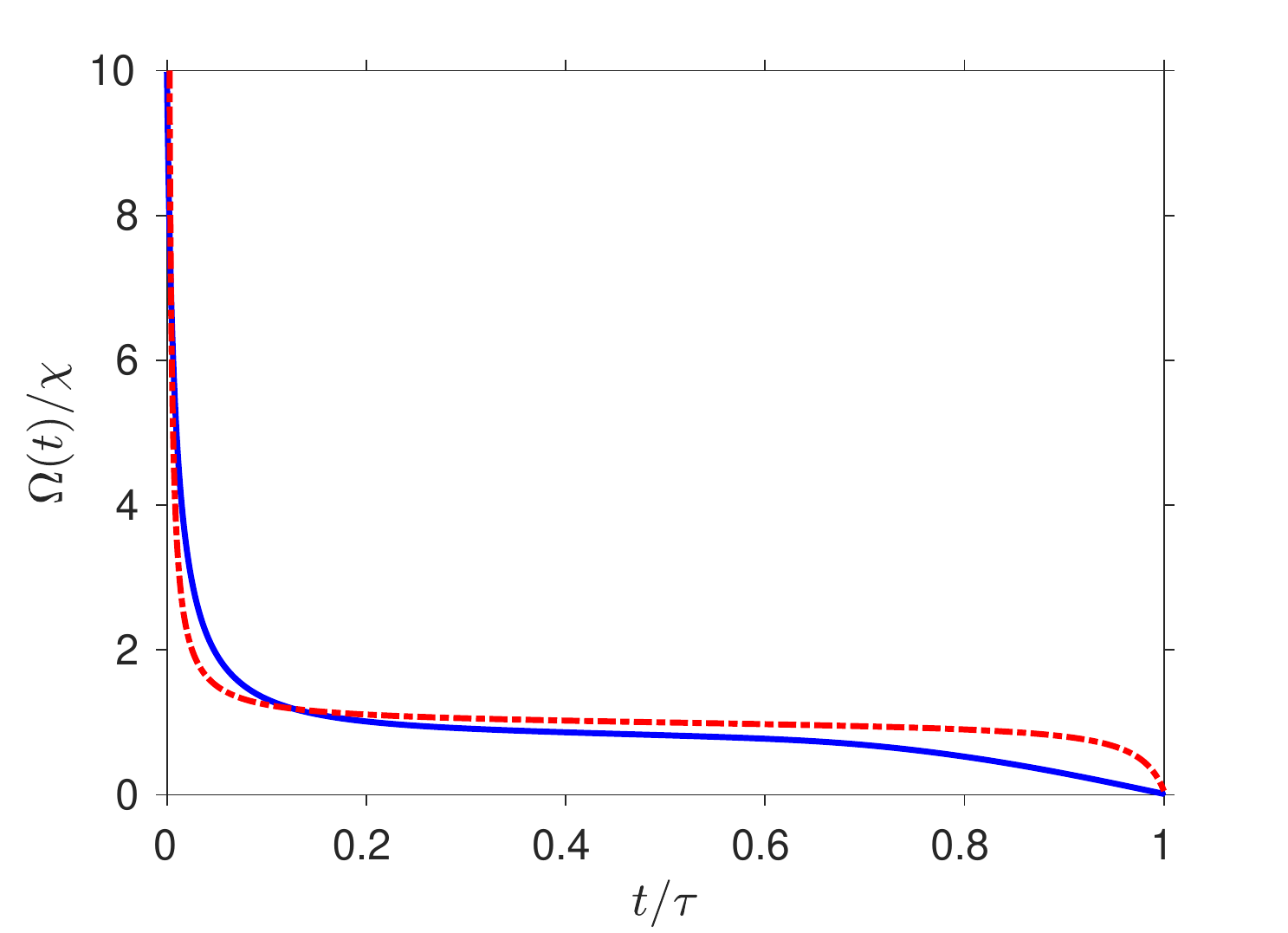}
 \caption{LAA sequence to ramp down transverse field $\Omega(t)$ for the LMG (solid blue) and TFI (dot-dashed red) models. Examples are for: (LMG) $N = 50$, $\chi\tau = 10$, and $(\Omega(0),\Omega(\tau)) = (10,10^{-2})$, and (TFI) $N = 20$, $\chi\tau = 10$, and $(\Omega(0),\Omega(\tau)) = (10^2,10^{-2})$.}
 \label{fig:LAAramp}
\end{figure}

The LAA ramp sequence entails varying the rate of change $d\Omega/dt$ such that diabatic excitations are created uniformly throughout the ramp. In principle, this is more efficient than, say, a linear ramp with fixed $d\Omega/dt$, for which the performance is greatly determined by excitations generated when the energy gap $\Delta$ between ground and excited states is smallest. A LAA ramp can be constructed via
\begin{eqnarray}
 \frac{d\Omega}{dt} = \frac{\Delta(t)^2}{\gamma} , \\
 \gamma = \frac{\tau}{\int_0^{\Omega(\tau)} \frac{d\Omega}{\Delta(B)^2}}
\end{eqnarray}
where $\Delta(t)$ [$\Delta(\Omega)$] is the energy gap of the instantaneous Hamiltonian (field strength). Example LAA ramps for the Lipkin-Meshkov-Glick (LMG) and $1$D transverse-field Ising (TFI) models are shown in Fig.~\ref{fig:LAAramp}.

\section{Pseudo-echo protocol for obtaining MQC spectrum}
In the main text we outline a pair of dynamical protocols [see also Fig.~3(a)] to obtain the MQC spectrum of the ground state of both TFI and LMG models for a given $\Omega/\chi$ through an echo or pseudo-echo sequence. We argued that if the ramp dynamics which underly these protocols is sufficiently adiabatic then both sequences are equivalent, i.e., the MQC spectrum and thus related fidelity out-of-time-order correlations from which it is constructed can be obtained without time-reversal of coherent dynamics. In this section we expand upon our reasoning for this connection and provide further justification with numerical simulations of the LMG and TFI models.

\subsection{Connection between ideal and pseudo-echo protocols}
Explicitly, the MQC spectrum of the ground state of both TFI and LMG models for a given $\Omega/\chi$ can be obtained via an echo sequence. This consists of i) initialization of the spins in the paramagnetic ground state $\hat{\rho}_0 = \vert (N/2)_x \rangle \langle (N/2)_x \vert$ at large initial field $\Omega(0) \gg \chi$, ii) an adiabatic ramp of duration $\tau$ wherein the transverse field is slowly quenched to a final value $\Omega(\tau)$ with $\chi$ held fixed, iii) a global rotation $\propto e^{-i\hat{\phi}\hat{S}_x}$, and iv) a many-body echo of the initial ramp wherein the sign of the Hamiltonian is flipped and the time-dependence of Hamiltonian parameters is reversed, such that $\hat{H}(t) = -\hat{H}(2\tau - t)$ for $t > \tau$. A measurement of the overlap of the resulting state $\hat{\rho}_f^{\phi}$ with respect to $\hat{\rho}_0$ for many $\phi$ then gives access to the MQC spectrum: 
\begin{eqnarray}
 \mathrm{Tr}\left[ \hat{\rho}_0 \hat{\rho}^{\phi}_f \right] & \equiv & \mathrm{Tr}\left[ \hat{\rho_0} \hat{U}^{\dagger}(\tau)e^{-i\phi\hat{S}_x} \hat{U}(\tau) \hat{\rho_0} \hat{U}^{\dagger}(\tau)e^{i\phi\hat{S}_x} \hat{U}(\tau) \right] , \notag \\
 & & = \mathrm{Tr}\left[ \rho(\tau)\hat{\rho}^{\phi}(\tau)\right] , \notag \\
 & & = \sum_m I^{\hat{S}_x}_m(\hat{\rho}(\tau)) e^{-im\phi} . \label{eqn:IdealEchoMQC}
\end{eqnarray}
Here, we have used the cyclic property of the trace and defined $\hat{\rho}(\tau) = \hat{U}(\tau) \hat{\rho_0} \hat{U}^{\dagger}(\tau)$ and $\hat{\rho}^{\phi}(\tau) = e^{-i\phi\hat{S}_x}\hat{\rho}(\tau)e^{i\phi\hat{S}_x}$ where $\hat{U}(\tau)$ is a unitary operator describing the ramp evolution. This many-body echo sequence, combined with the measurement $\mathrm{Tr}\left[ \hat{\rho}_0 \hat{\rho}^{\phi}_f \right]$, is a type of fidelity out-of-time-order correlation (FOTOC), which was studied in depth in Ref.~\cite{RLS_2019scrambling}.

Having recapped the ideal echo protocol which allows one to explicitly construct the MQC spectrum of the state $\hat{\rho}(\tau)$ (and thus in principle $\hat{\rho}_{\mathrm{GS}}$), we now discuss in more detail our statement from the main paper wherein we report that the MQC spectrum can also be obtained in the absence of true time-reversal. In particular, we consider the scenario where in step iv) of the ideal echo protocol the sign of the Hamiltonian is not flipped: $\hat{H}(t) = \hat{H}(2\tau - t)$ for $t > \tau$. We term this a \emph{pseudo-echo}. In the following we will i) investigate what information can be extracted when one measures the overlap of the final state of the pseudo-echo $\hat{\rho}^{\prime,\phi}_f$ with $\hat{\rho}_0$, and ii) formally justify our argument that this measurement precisely gives the MQC spectrum of the ground state in the limit that the ramping dynamics are adiabatic. 

Formally, the overlap with $\hat{\rho}_0$ after the pseudo-echo can be expanded as
\begin{eqnarray}
 \mathrm{Tr}\left[ \hat{\rho}_0 \hat{\rho}^{\prime\phi}_f \right] & \equiv & \mathrm{Tr}\left[ \hat{\rho_0} \hat{U}^{\prime}(\tau)e^{-i\phi\hat{S}_x} \hat{U}(\tau) \hat{\rho_0} \hat{U}^{\dagger}(\tau)e^{i\phi\hat{S}_x} \hat{U}^{\prime\dagger}(\tau) \right] , \notag \\
 & = & \mathrm{Tr}\left[ \hat{\rho}^{\prime}(\tau) \hat{\rho}^{\phi}(\tau) \right], \notag \\
 & \equiv & \sum_m \tilde{I}^{\hat{S}_x}_m(\hat{\rho}^{\prime}(\tau),\hat{\rho}(\tau)) e^{-im\phi} . \label{eqn:PseudoEchoMQC}
\end{eqnarray}
Here, the second ramp in the pseudo-echo is described by the unitary operator $\hat{U}^{\prime}(\tau)$, $\hat{\rho}^{\prime}(\tau) = \hat{U}^{\prime\dagger}(\tau) \hat{\rho}_0 \hat{U}^{\prime}$ and we have defined effective intensities:
\begin{equation}
 \tilde{I}^{\hat{S}_x}_m(\hat{\rho}^{\prime}(\tau),\hat{\rho}(\tau)) \equiv \mathrm{Tr}\left[ \hat{\rho}^{\prime}_{-m}(\tau) \hat{\rho}_m(\tau) \right]
\end{equation}
If $\hat{U}^{\prime}(\tau) = \hat{U}^{\dagger}(\tau)$ then we would recover the connection to the MQC spectrum as per Eq.~(\ref{eqn:IdealEchoMQC}), i.e. $\tilde{I}^{\hat{S}_x}_m(\hat{\rho}^{\prime}(\tau),\hat{\rho}(\tau)) \to I^{\hat{S}_x}_m(\hat{\rho}(\tau))$.

A pseudo-echo does not realize $\hat{U}^{\prime}(\tau) = \hat{U}^{\dagger}(\tau)$ due to the difference in sign of the Hamiltonian $\hat{H}(t)$. However, if the ramping protocol is sufficiently adiabatic, i.e. $\tau \to \infty$ for a finite system, then it is possible to mimic this relation. To understand this claim, consider a ramp $\hat{U}(\tau\to\infty)$ which \emph{adiabatically} transforms the initial ground state $\vert \psi_{\mathrm{GS}}(0) \rangle$ of $\hat{H}(0)$ into $\vert \psi_{\mathrm{GS}}(\tau) \rangle$ for $\hat{H}(\tau)$. Then, implement a subsequent adiabatic ramp back to $\hat{H}(2\tau) = \hat{H}(0)$ described by $\hat{U}^{\prime}(\tau\to\infty)$. Clearly, if the last ramp is adiabatic then we expect $\vert \psi_{\mathrm{GS}}(2\tau) \rangle = \hat{U}^{\prime}(\tau\to\infty)\vert \psi_{\mathrm{GS}}(\tau) \rangle \propto \vert \psi_{\mathrm{GS}}(0) \rangle$ up to an irrelevant global phase. This suggests that $\hat{U}^{\prime}(\tau\to\infty) \simeq \hat{U}^{\dagger}(\tau\to\infty)$ up to some irrelevant phase. Thus, \emph{if the ramps are adiabatic} then $\tilde{I}^{\hat{S}_x}_m(\hat{\rho}^{\prime}(\tau),\hat{\rho}(\tau)) \to I^{\hat{S}_x}_m(\hat{\rho}(\tau))$ and hence the pseudo-echo protocol can be used to obtain the MQC spectrum.

\subsection{Numerical simulations}
\begin{figure}[tb!]
 \includegraphics[width=8cm]{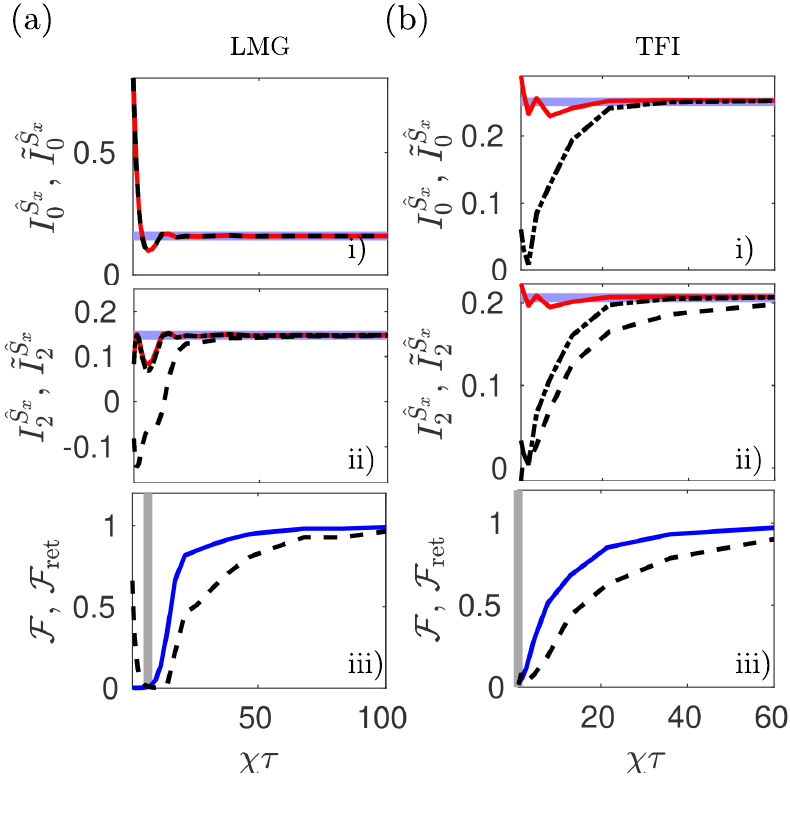}
 \caption{Evaluation of performance of the pseudo-echo in obtaining the MQC spectrum, as compared to the ideal echo protocol for (a) LMG and (b) TFI models. Intensities are computed for a quench to $\Omega(\tau)/\chi = 10^{-2}$ from $\Omega(0)/\chi = 10^{1}$ (LMG) and $\Omega(0)/\chi = 10^{2}$ (TFI). Panels (i)-(ii) Comparison of ideal $I^{\hat{S}_x}_m$ of exact ground state (thick blue), adiabatic ramp with ideal echo (red) and $\tilde{I}^{\hat{S}_x}_m$ obtained from a pseudo-echo (dashed black). We also plot $\vert \tilde{I}^{\hat{S}_x}_m \vert$ (dot-dashed black) which shows improved agreement. Note that in (a), $\tilde{I}^{\hat{S}_x}_0=I^{\hat{S}_x}_0$ due to the symmetries of the LMG model (see text). Panel (iii) Correlation of return fidelity $\mathcal{F}_{\mathrm{ret}}$ (dashed black) with fidelity $\mathcal{F}$ (solid blue) of prepared state to the actual ground state (for $\Omega(\tau) = 10^{-2}$) as a function of ramp duration $\tau$. Vertical grey line indicates characteristic interaction timescale $\chi\tau = \sqrt{N}$ (LMG) and $\chi\tau = 1$ (TFI) (see text).}
 \label{fig:PseudoEcho}
\end{figure}

\begin{figure*}[tb!]
 \includegraphics[width=16cm]{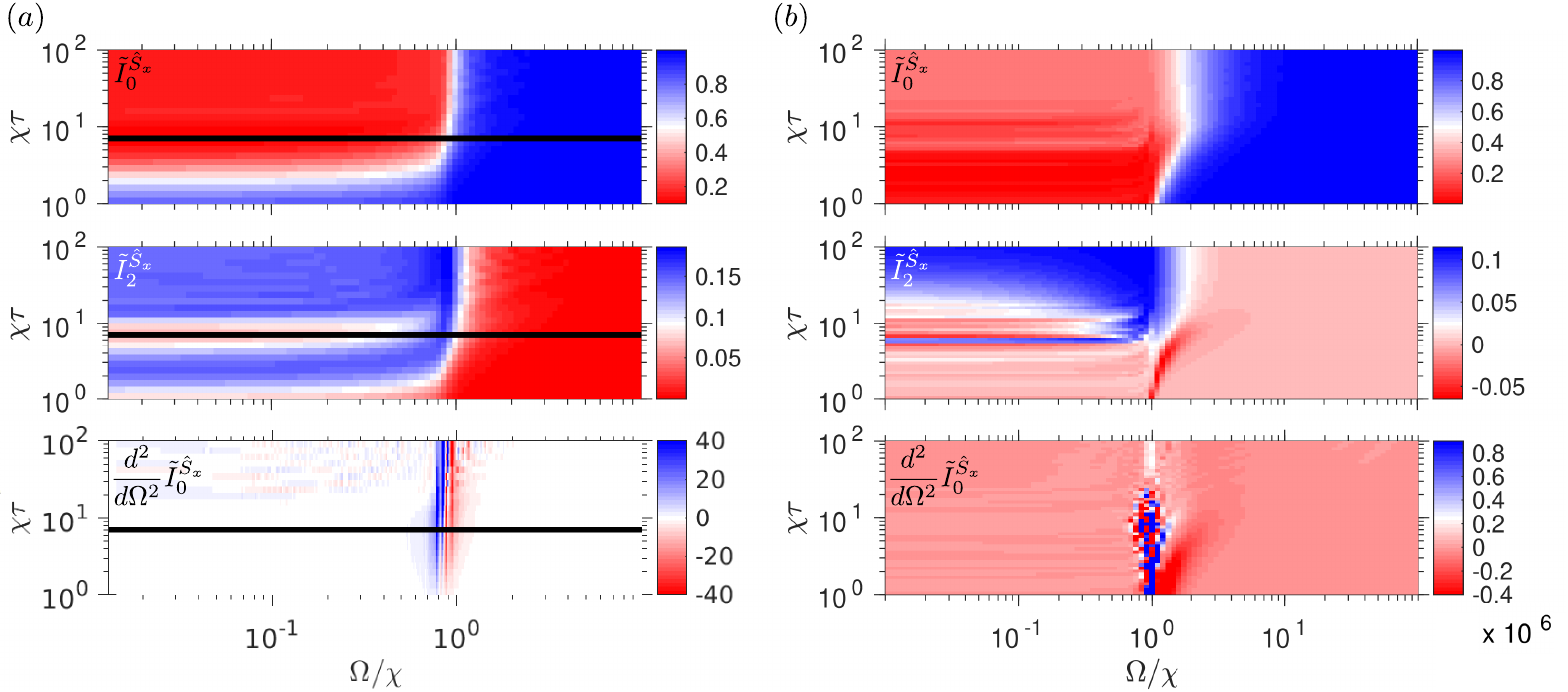}
 \caption{\Rev{Robustness of QPT signatures in the effective MQC intensities $\tilde{I}^{\hat{S}_x}_{0,2}$ and derivative $\frac{d^2}{d\Omega^2}\tilde{I}^{\hat{S}_x}_0$ as a function of total ramp duration $\tau$ for (a) LMG and (b) TFI models. For the LMG model we plot the characteristic interaction timescale $\chi\tau = \sqrt{N}$ for reference as a horizontal black line. The equivalent characteristic timescale for the TFI model is $\chi \tau = 1$, which coincides with our shortest ramps. Note that for the plots of the derivative the lower limit of the colormap is saturated for some features (particularly the noisy structure around $\Omega/\chi \approx 1$ for $\chi\tau \lesssim 10$ for the TFI model). All results are from numerical solution of the time-dependent Schr\"{o}dinger equation with $N = 50$ (LMG) and $N=20$ (TFI). For both models, $\frac{d^2}{d\Omega^2}\tilde{I}^{\hat{S}_x}_0$ is computed numerically using a step size of $\delta(\Omega/\chi) = 10^{-4}$.}}
 \label{fig:PseudoEcho_Features}
\end{figure*}

We justify our use of the pseudo-echo protocol with numerical simulations comparing against the ideal many-body echo for both the LMG and TFI models, as summarized in Figs.~\ref{fig:PseudoEcho} and \ref{fig:PseudoEcho_Features}. In all plots we vary the total ramp duration $\tau$ for a small system (LMG $N=50$, TFI $N=20$) to evaluate the performance of the pseudo-echo protocol as a function of the adiabaticity of the ramp. In Fig.~\ref{fig:PseudoEcho} we compare the final intensities $\tilde{I}^{\hat{S}_x}_m$ to those of an ideal echo, ${I}^{\hat{S}_x}_m$, obtained at the end of ramps from $\Omega/\chi = 10^1$ (LMG) or $\Omega/\chi = 10^2$ (TFI) to a final $\Omega/\chi = 10^{-2}$. We observe that, typically, the $\tilde{I}^{\hat{S}_x}_m$ rapidly approach the true intensities $I^{\hat{S}_x}_m$, with increasing $\tau$. An exception is the $m=0$ intensity for the LMG model, which we find precisely agrees $\tilde{I}^{\hat{S}_x}_0=I^{\hat{S}_x}_0$; this peculiarity is due to a symmetry possessed by both the LMG and TFI models in addition to the lack of degeneracy in the LMG Hamiltonian, and is discussed in the following subsection. The associated adiabaticity of the ramp, quantified approximately through the fidelity $\mathcal{F}$ of the dynamically prepared state to the equivalent ground state at $\Omega(\tau)/\chi$, demonstrates that when $\mathcal{F} \gtrsim 0.7$ excellent agreement is found between the two MQC intensities. For completeness, we also plot the return fidelity $\mathcal{F}_{\mathrm{ret}}$, which was introduced in the main text. This quantity is obtained by measuring the overlap of the final and initial states of the pseudo-echo protocol in the absence of any perturbation ($\phi = 0$). The return fidelity serves as a useful way to operationally determine the adiabaticity of the ramps. In this context, one subtlety which must be dealt with properly is that $\mathcal{F}_{\mathrm{ret}}$ is only useful as a proxy for $\mathcal{F}$ when $\chi\tau$ is greater than the characteristic timescale of interactions in the model: $\chi\tau \gtrsim \sqrt{N}$ and $\chi\tau \gtrsim 1$ for the LMG and TFI models, respectively. This is because for shorter ramps no appreciable dynamics can occur and thus, redundantly, $\mathcal{F}_{\mathrm{ret}} \sim 1$.

\Rev{Similarly, in Fig.~\ref{fig:PseudoEcho_Features} we assess how robust the signatures of the QPT in the individual intensities for $m=0,2$ are for the pseudo-echo protocol. For the LMG model we observe that the abrupt transition in both $\tilde{I}^{\hat{S}_x}_0$ and $\tilde{I}^{\hat{S}_x}_2$ is robust across a range of ramp durations $\tau$. Significant deviations only emerge in the limit of very short ramps $\chi\tau \lesssim \sqrt{N}$ as previously discussed. Moreover, we also find the peak in the derivative $\frac{d^2}{d\Omega^2}\tilde{I}^{\hat{S}_x}_0$ (blue region in lower left panel of Fig.~\ref{fig:PseudoEcho_Features}) is clear even for fast ramps, and shows no systematic offset from the effective critical point $(\Omega/\chi)_*$ as a function of $\chi\tau$. Rather, the peak of the feature appears to simply vanish as $\chi\tau \to 0$. For the TFI model, we also observe a robustness of the QPT signatures in the MQC intensities for a range of ramp durations, though the signature in $\tilde{I}^{\hat{S}_x}_0$ is generally more stable for a larger range of ramp times than that in $\tilde{I}^{\hat{S}_x}_2$. Deviations in both quantities begin to appear for shorter ramp times, though for $\chi\tau$ larger than the characteristic interaction timescale we nevertheless observe that $\tilde{I}^{\hat{S}_x}_0$ develops sharp features at the critical point. The peak in the derivative $\frac{d^2}{d\Omega^2}\tilde{I}^{\hat{S}_x}_0$ is robust in that it shows a strong signature of the QPT for even short ramps. However, it is dominated by fine structure/noise for $\chi\tau \lesssim 30$ before saturating to the expected ground state prediction (white region near $\chi/\Omega \approx 1$ in lower right panel of Fig.~\ref{fig:PseudoEcho_Features}).}

\subsection{Exact connections between MQC intensities due to symmetries of the LMG and TFI model}
In the previous subsection, we have established that, excluding the case of adiabatic dynamics, there appears to be no exact correspondence between the MQC intensities obtained from the pseudo-echo and ideal many-body echo protocols in general. However, as hinted at by the results of Fig.~(\ref{fig:PseudoEcho}), in the special case where the dynamics is described by the LMG or TFI models investigated here, and with certain further assumptions, it is actually possible to directly relate the effective MQC intensities $\tilde{I}^{\hat{S}_x}_m$ and the true MQC intensities $I^{\hat{S}_x}_m$. This is specifically related to the fact that the generic Hamiltonian
\begin{equation}
 \hat{H} = -\frac{1}{2\mathcal{C}} \sum_{i < j} \chi_{ij}\hat{\sigma}^z_i\hat{\sigma}^z_{j} - \frac{\Omega}{2}\sum_i \hat{\sigma}^x_i , \label{eqn:Hgeneric}
\end{equation}
which describes both the TFI ($\chi_{ij} = \chi\delta_{i,j-1}$, so that $\mathcal{C} = 1$) and LMG ($\chi_{ij} = \chi$, so that $\mathcal{C} = N$) models, is real symmetric: $\hat{H}^* = \hat{H}$ where ${}^*$ denotes complex conjugation. In the following, we will explicitly prove this results in a connection between $\tilde{I}^{\hat{S}_x}_m$ and $I^{\hat{S}_x}_m$ for the LMG model, including the result that $\tilde{I}^{\hat{S}_x}_0 = I^{\hat{S}_x}_0$, and then comment on the implications for the TFI model.

\subsubsection{Connection for ramping protocol}
We consider a toy model of the ramping sequences described by $\hat{U}(\tau)$ and $\hat{U}^{\prime\dagger}(\tau)$ constructed by discretizing them into $j=1,2,...,n$ steps of duration $\Delta\tau = \tau/n$ each described by the instantaneous LMG Hamiltonian (Eq.~(\ref{eqn:Hgeneric}) with $\mathcal{C} = N$ and $\chi_{ij} = \chi$) $\hat{H}_n \equiv \hat{H}\left((n-1/2)\Delta\tau\right)$: 
\begin{multline}
 \hat{U}(\tau) = \lim_{n\to\infty} \left[ e^{-i\hat{H}_n\Delta\tau} e^{-i\hat{H}_{n-1}\Delta\tau} ... e^{-i\hat{H}_2\Delta\tau} e^{-i\hat{H}_1\Delta\tau} \right] , \\
 \hat{U}^{\prime}(\tau) = \lim_{n\to\infty} \left[ e^{-i\hat{H}_1\Delta\tau} e^{-i\hat{H}_{2}\Delta\tau} ... e^{-i\hat{H}_{n-1}\Delta\tau} e^{-i\hat{H}_n\Delta\tau} \right] .
\end{multline}
As will become clear, it is also useful to note that in this model
\begin{equation}
 \hat{U}^{\prime\dagger}(\tau) = \lim_{n\to\infty} \left[ e^{i\hat{H}_n\Delta\tau} e^{i\hat{H}_{n-1}\Delta\tau} ... e^{i\hat{H}_2\Delta\tau} e^{i\hat{H}_1\Delta\tau} \right] .
\end{equation}
Trivially, for an initial pure state $\vert \psi(0) \rangle$ we have that $\hat{\rho}(\tau) = \vert \psi(\tau) \rangle \langle \psi(\tau) \vert$ with $\vert \psi(\tau) \rangle = \hat{U}(\tau) \vert \psi(0) \rangle$, and similarly $\hat{\rho}^{\prime}(\tau) = \vert \psi^{\prime}(\tau) \rangle \langle \psi^{\prime}(\tau) \vert$ with $\vert \psi^{\prime}(\tau) \rangle = \hat{U}^{\prime\dagger}(\tau) \vert \psi(0) \rangle$.

A consequence of $\hat{H}_j$ being real symmetric is that any associated evolution is complex symmetric: i.e., $\left(e^{-i\hat{H}_j\Delta\tau} \right)^{\dagger} \equiv \left(e^{-i\hat{H}_j\Delta\tau} \right)^* = e^{i\hat{H}_j\Delta\tau}$. If we assume that the initial state can be expanded in the fully symmetric Dicke basis as $\vert \psi(0) \rangle = \sum_m c_m(0) \vert m_x \rangle$ with $c_m(0) \in \mathbb{R}$ then the fact that the unitary operator is complex symmetric leads us to the useful result $\vert \psi^{\prime}(\tau) \rangle = \sum_m c^*_m(\tau) \vert m_x \rangle$ and $\vert \psi(\tau) \rangle = \sum_m c_m(\tau) \vert m_x \rangle$. The fact that the two states $\vert \psi^{\prime}(\tau) \rangle$ and $\vert \psi(\tau) \rangle$ share the same coefficients up to conjugation allows us to directly relate the MQC intensities:
\begin{eqnarray}
 I^{\hat{S}_x}_m(\hat{\rho}) = \sum_n \vert c_n \vert^2 \vert c_{n+m} \vert^2 , \label{eqn:Icoeff} \\
 \tilde{I}^{\hat{S}_x}_m(\hat{\rho},\hat{\rho}^{\prime}) = \sum_n c_n^2 c_{n+m}^{*2} . \label{eqn:tildeIcoeff}
\end{eqnarray}
From these equations, we can immediately deduce $\tilde{I}^{\hat{S}_x}_0 = I^{\hat{S}_x}_0$. Moreover, the form of Eqs.~(\ref{eqn:Icoeff}) and (\ref{eqn:tildeIcoeff}) allow us to establish a bound on curvature $\sum_m m^2\vert I^{\hat{S}_x}_m \vert \geq \sum_m m^2\vert \tilde{I}^{\hat{S}_x}_m \vert$. This inequality has a subsequent practical application as a bound on the quantum Fisher information \cite{Braunstein1994,Martin_2018,Garrahan_2019} $\mathcal{F}_Q \geq 2\sum_m m^2\vert \tilde{I}^{\hat{S}_x}_m \vert$. 

The results (\ref{eqn:Icoeff}) and (\ref{eqn:tildeIcoeff}) explain why we explicitly observe $\tilde{I}^{\hat{S}_x}_0 = I^{\hat{S}_x}_0$ in Fig.~\ref{fig:PseudoEcho} for the LMG model. Morever, as also evidenced in Fig.~\ref{fig:PseudoEcho}, we find that $\vert \tilde{I}^{\hat{S}_x}_m \vert$ typically follows ${I}^{\hat{S}_x}_m$ more closely, particularly for ramps which are not adiabatic. In this context, we always plot $\vert \tilde{I}^{\hat{S}_x}_m \vert$ in Fig.~3(b)-(c) of the main text.

\subsubsection{Extension to TFI model}
These findings for the LMG model can be extended to the TFI case after appropriately accounting for the degeneracy of states with total spin projection $m_x$. Specifically, if we expand $\vert \psi(\tau) \rangle = \sum_{m,\alpha} c_{m,\alpha} \vert m_x,\alpha\rangle$ where $\hat{S}_x\vert m_x,\alpha \rangle = m \vert m_x,\alpha \rangle$ and $\alpha$ labels each degenerate state, then for the TFI model we have the more general results
\begin{eqnarray}
 I^{\hat{S}_x}_m(\hat{\rho}) = \sum_{n,\alpha,\alpha^{\prime}} \vert c_{n,\alpha} \vert^2 \vert c_{n+m,\alpha^{\prime}} \vert^2 , \label{eqn:TFIIcoeff} \\
 \tilde{I}^{\hat{S}_x}_m(\hat{\rho},\hat{\rho}^{\prime}) = \sum_{n,\alpha,\alpha^{\prime}} c_{n,\alpha}^2 c_{n+m,\alpha^{\prime}}^{*2} . \label{eqn:TFItildeIcoeff}
\end{eqnarray}
Thus, for the TFI model and for more general models described by the Hamiltonian (\ref{eqn:Hgeneric}) which probe the full $2^N$ Hilbert space, Eqs.~(\ref{eqn:TFIIcoeff}) and (\ref{eqn:TFItildeIcoeff}) imply that $\tilde{I}^{\hat{S}_x}_0 \neq I^{\hat{S}_x}_0$ unlike the LMG model, though the aforementioned bound on the curvature still holds.

\section{Exact solutions for MQC spectrum}
The ground state and associated MQC spectrum can be solved analytically for both the LMG and TFI models. In the following we outline the details of the exact solutions and analytic expressions for the MQC intensities. The results of these calculations are plotted in Figs.~1 and 2 of the main text.

\subsection{TFI model}
We first focus on the transverse-field Ising model in $1$D. Assuming periodic conditions (PBC), the TFI model is described by the Hamiltonian 
\begin{align} 
\hat{H}_{\textrm{TFI}} = -\sum_{i=1}^N \hat{\sigma}^z_i\hat{\sigma}^z_{i+1} - g\sum_{i=1}^N \hat{\sigma}^x_i ,
\end{align}
for Pauli operators $\hat{\sigma}^{\alpha}_i$, $\alpha = x,y,z$, and where we define the site indices modulo the number of lattice sites, $N$. This Hamiltonian is equivalent to Eq.~(1) of the main text (up to an overall re-scaling) for $\chi_{ij} = \chi\delta_{i,j-1}$ so that $\mathcal{C} = 1$, where we imposed periodic boundary conditions and defined $g=\Omega /\chi$ for convenience.

\subsubsection{Exact ground state}
To find the ground state, we follow the standard procedure of first rotating about the y-axis by an angle $\pi/2$ so that $\hat{\sigma}^x_i \rightarrow -\hat{\sigma}^z_i$, $\hat{\sigma}^z_i \rightarrow \hat{\sigma}^x_i$, and then applying a Jordan-Wigner transformation \cite{Pfeuty_TFI_1970}
\begin{align}
 \hat{\sigma}^+_i = \hat{c}^\dagger_ie^{i\pi\sum_{j<i}\hat{n}_j},\qquad \hat{\sigma}^z_i = 2\hat{n}_i - 1
\end{align}
with fermionic creation operators $\hat{c}^\dagger_i$ and number operators $\hat{n}_i \equiv \hat{c}^\dagger_i\hat{c}_i$. Thus, our Hamiltonian becomes
\begin{multline}
 \hat{H}_{\textrm{JW}} = -\sum_{i=1}^{N-1} \left(\hat{c}^\dagger_i - \hat{c}_{i}\right)\left(\hat{c}^\dagger_{i+1} + \hat{c}_{i+1}\right) \\
 + \hat{P}\left(\hat{c}^\dagger_N - \hat{c}_{N}\right)\left(\hat{c}^\dagger_1 + \hat{c}_{1}\right) + g\sum_{i=1}^N \left(2\hat{n}_i - 1\right)
\end{multline}
where $\hat{P} = e^{i\pi\sum_j \hat{n}_j}$ is the fermion number parity operator for the system. For states of odd fermion parity, this Hamiltonian is translationally invariant and possesses PBC; however, for states of even fermion parity, the Hamiltonian instead possesses anti-periodic boundary conditions (APBC). For both cases, we may write $\hat{H}_{\textrm{JW}}$ in terms of quasimomentum modes as
\begin{multline}
\hat{H}_{\textrm{JW}} = -\sum_{k\in \textrm{B.Z.}}\big[2\left(\cos{k} - g\right)\hat{n}_k\\
\qquad\qquad- i\sin{k}\left(\hat{c}^\dagger_k\hat{c}^\dagger_{-k} - h.c.\right) + g\big]
\end{multline}
for fermion number operator $\hat{n}_k = \hat{c}^\dagger_k\hat{c}_k$, where we have substituted $\hat{c}_j = \frac{1}{\sqrt{N}}\sum_{k\in \textrm{B.Z.}}\hat{c}_ke^{-ijk}$
and assumed a lattice constant $a=1$ for simplicity. For states with odd fermion parity, the sum over the first Brillouin zone (B.Z.) runs over quasimomentum modes $k=2\pi n/N$ with integer $n=0,1,...,N-1$; however, for states with even fermion parity, this sum instead runs over quasimomentum modes $k=2\pi n/N$ with half-integer $n=1/2,3/2,...,N-1/2$. For a description of the original physical model, since we have conservation of fermion number parity we can therefore neglect the set of unphysical states corresponding to even (odd) fermion parity states of integer (half-integer) quasimomentum modes.

We now apply a Bogoliubov transformation $\hat{c}_k = u_k\hat{\gamma}_k + iv_k\hat{\gamma}_{-k}^\dagger$ where $u_k = \cos{(\theta_k/2)}$ and $v_k = \sin{(\theta_k/2)}$ for some angle $\theta_k$. To conserve particle number, we choose $\tan{\theta_k} = \sin{k}/(\cos{k}-g)$, leaving us with
\begin{align}
 \hat{H}_{\textrm{Bog}} = \sum_{k\in\textrm{B.Z.}} \varepsilon_k \hat{\gamma}^\dagger_k\hat{\gamma}_k + E_0
\end{align}
for mode energy
\begin{align}
 \varepsilon_k &= 2\sqrt{g^2 - 2g\cos{k} + 1}
\end{align}
and constant offset $E_0$ which we may neglect.

The ground state of $\hat{H}_{\textrm{Bog}}$ is the vacuum with respect to $\hat{\gamma}_k$, denoted by $\ket{\psi_{\mathrm{GS}}}_{\gamma}$. This gap vanishes at $g = 1$ in the thermodynamic limit, indicating a transition between a quantum ferromagnetic phase for $g < 1$ and a quantum paramagnetic phase for $g > 1$. In terms of our original quasimomentum modes, it is straightforward to show
\begin{align}
 \ket{\psi_{\mathrm{GS}}}_{\gamma} \propto \prod_{0\leq k < \pi} \left[1 + i\tan{\left(\theta_k/2\right)}\hat{c}^\dagger_k\hat{c}^\dagger_{-k}\right]\ket{0}
\end{align}
where $\ket{0}$ is the vacuum with respect to $\hat{c}_k$ and $k$ runs over only half the Brillouin zone to avoid over-counting. Since this state has even parity, the ground state of our original spin model, $\ket{\psi_{\mathrm{GS}}}$, only includes half-integer quasimomentum modes, and thus corresponds exclusively to the ground state of the APBC fermion model. Properly normalizing this state, we arrive at
\begin{align}
 \ket{\psi_{\mathrm{GS}}} = \prod_{k} \left[\cos{\left(\theta_k/2\right)} + i\sin{\left(\theta_k/2\right)}\hat{c}^\dagger_k\hat{c}^\dagger_{-k}\right]\ket{0} \label{eqn:TFIgs}
\end{align}
where the product index $k$ will now be understood as running over only half-integer quasimomentum modes with $0\leq k <\pi$.

\subsubsection{FOTOC}
To obtain the required MQC spectrum from the ground state Eq.~(\ref{eqn:TFIgs}) we must first compute the associated FOTOC. Specifically we compute $F_{\phi} = \mathrm{Tr}\left[\hat{\rho}_{\mathrm{GS}}\hat{\rho}^{\phi}_{\mathrm{GS}} \right]$ where $\hat{\rho}_{\mathrm{GS}} = \ket{\psi_{\mathrm{GS}}} \bra{\psi_{\mathrm{GS}}}$ and $\hat{\rho}_{\mathrm{GS}}^{\phi} = e^{-i\phi\hat{S}_x}\hat{\rho}_{\mathrm{GS}}e^{i\phi\hat{S}_x}$. The latter is simplified via relating
\begin{align}
 \hat{S}_x = \frac{1}{2}\sum_i\hat{\sigma}_i^x \rightarrow \sum_k \hat{n}_k ,
\end{align}
and we thus find
\begin{align}
\begin{split}
 F_\phi &= \big|\braket{\psi_{\mathrm{GS}}|e^{-i\phi\hat{S}_x}|\psi_{\mathrm{GS}}}\big|^2\\
 &= \prod_{k} \left[1 - \sin^2{(\phi)}f(k,g)\right]\label{eq:TFI_FOTOC_prod}
 \end{split}
\end{align}
where $f(k,g) = 1/[1 + (\cos{k} - g)^2/\sin^2{k}]$.

To find an exact closed-form expression for Eq.~(\ref{eq:TFI_FOTOC_prod}) for arbitrary system size $N$, we first have that
\begin{align}
 1 &- \sin^2(\phi)f(k,g) \label{eq:TFI_factor}\\
 \begin{split}&=\frac{\sin^2(\phi)\sin^2(k/2)}{g}\\
 &\qquad\times\frac{\left[1-X_{+,\phi}(g)/\sin^2(k/2)\right]\left[1-X_{-,\phi}(g)/\sin^2(k/2)\right]}{\left[1-(-\sinh^2((\ln{g})/2))/\sin^2(k/2)\right]}\end{split}\label{eq:factorized_kTerm}
\end{align}
for functions 
\begin{multline} 
 X_{\pm,\phi}(g) = \frac{1}{2}\Bigg\{1 - \frac{g}{\sin^2(\phi)}\\\times\Bigg[1
 \mp \sqrt{1-\sin^2(\phi)\left(1+\frac{1}{g^2}\right) + 
 \frac{\sin^4(\phi)}{g^2}}\Bigg]\Bigg\}, \label{eq:Xpm}
\end{multline}
which results from writing Eq.~\eqref{eq:TFI_factor} as a single rational expression and then factorizing the numerator, which is a quadratic in $\sin^2(k/2)$. Next, we use the following relations \cite{gradshteyn2007},
\begin{gather}
 \prod_{n=1}^{\lfloor N/2 \rfloor}\sin^2\left(\frac{(2n-1)\pi}{2N}\right) = \frac{1}{2^{N-1}},\\
\prod_{n=1}^{\lfloor N/2 \rfloor}\left(1-\frac{\sin^2z}{\sin^2\left(\frac{(2n-1)\pi}{2N}\right)}\right) 
= \left\{\begin{aligned} &\cos(Nz),\;&&N\,\textrm{even}\\
&\frac{\cos(Nz)}{\cos z},&&N\,\textrm{odd}
\end{aligned}\right.,
\end{gather}
to evaluate the quasimomenta product in Eq.~\eqref{eq:TFI_FOTOC_prod} using the expression in Eq.~\eqref{eq:factorized_kTerm}. Assuming that $N$ is even, we thus obtain
\begin{gather} 
\begin{split}
 F_{\phi} = \frac{4}{1+g^N}\left(\frac{\sin(\phi)}{2}\right)^N\\
 \times\cos(N\arcsin{\sqrt{X_{+,\phi}(g)}})\\
 \times\cos(N\arcsin{\sqrt{X_{-,\phi}(g)}}).
 \end{split}\label{eq:TFI_FOTOC_exact}
\end{gather}

For the case of odd $N$, the only difference is that Eq.~\eqref{eq:TFI_FOTOC_exact} is divided through by an additional factor of $\cos(\arcsin\sqrt{X_{+,\phi}(g)})\cos(\arcsin\sqrt{X_{-,\phi}(g)})$. We note in passing that for even $N$, we may alternatively express the cosine expressions in terms of the Chebyshev polynomials as $\cos(N\arcsin\sqrt{X_{\pm,\phi}(g)}) = T_{N/2}(1-2X_{\pm,\phi}(g))$, where $T_{\alpha}$ is the $\alpha$-th Chebyshev polynomial of the first kind.

It is also useful to obtain a more concise expression for Eq.~(\ref{eq:TFI_FOTOC_exact}) in the limit of large $N$, which is equivalent to assuming a continuum of quasimomentum modes. Exponentiating our product in Eq.~\eqref{eq:TFI_FOTOC_prod} and replacing the resulting sum with an integral, $\sum_k \rightarrow N\int^{\pi}_0 dk/(2\pi)$, then 
\begin{gather}
 F_{\phi} \simeq e^{-N\lambda_{\phi}(g)},\label{eq:FOTOC_cont}\\
 \lambda_{\phi}(g) = -\int_0^\pi \frac{dk}{2\pi}\, \ln{\left[1-\sin^2(\phi)f(k,g)\right]}. 
\end{gather}
Let us now define the quantity
\begin{align}
 A_n(g) \equiv \int_0^\pi \frac{dk}{2\pi}\, f(k,g)^n = \frac{a_n}{2}\begin{cases} \frac{1}{2^{2n}}, \quad g<1\\
 \frac{1}{(2g)^{2n}}, \quad g>1\end{cases}
\end{align}
for central binomial coefficient $a_n = \binom{2n}{n}$; we thus have that $A_n(g) = (a_n/2)A_1(g)^n$. To now evaluate $\lambda_{\phi}(g)$, we expand the logarithm in the integrand and perform the $k$ integral on each successive term, finding
\begin{align}
 \lambda_{\phi}(g) = \sum_{n=1}^\infty \frac{a_n}{2n}\left(\sin^2(\phi)A_1(g)\right)^n.
\end{align}
Introducing an auxiliary integral to write each term as
\begin{equation}
 \frac{a_n}{2n}\left(\sin^2(\phi)A_1(g)\right)^n = \lim_{\varepsilon\rightarrow 0} \int^{\sin^2(\phi)A_1(g)}_{\varepsilon/4} \frac{dx}{2x}\,a_n x^{n} ,
\end{equation}
we can use the fact that $(1-4x)^{-1/2}$ is the generating function for the central binomial coefficients, i.e. $\sum_{n=0}^\infty a_n x^n = (1-4x)^{-1/2}$, to eliminate the infinite sum and perform the integral over $x$, taking the limit $\varepsilon\rightarrow 0$ at the end to cancel singularities arising from the $n=0$ term. We thus obtain
\begin{equation}
 \lambda_{\phi}(g) = -\ln\left(\frac{1+\sqrt{1-4\sin^2(\phi)A_1(g)}}{2}\right), \label{eq:TFI_rate_func}
\end{equation}
and so
\begin{equation}
\begin{split}
 F_{\phi} \simeq \left\{\begin{aligned} &\left(\frac{1+|\cos(\phi)|}{2}\right)^N,\;&&g<1\\
&\left(\frac{1+\sqrt{1-\sin^2(\phi)/g^2}}{2}\right)^N,&&g>1\label{eq:TFI_FOTOC_cont}
\end{aligned}\right. .
\end{split}
\end{equation}

\subsubsection{MQC spectrum}
\Rev{The MQC intensities may be straightforwardly calculated via a Fourier transform of the exact FOTOC expressions in Eq.~\eqref{eq:TFI_FOTOC_prod} or Eq.~\eqref{eq:TFI_FOTOC_exact}. Numerically, we may generally obtain exact results for finite systems via a finite Fourier transform utilizing $2N$ points. Directly at the transition, however, we find that the MQC intensities possess a rather simple analytic form.} For $g = 1$, from Eq.~\eqref{eq:Xpm} we have $X_{+,\phi}(g=1) = 0$ and $X_{-,\phi}(g=1) = -\cot^2(\phi)$, and from this we have
\begin{align}
 F_{\phi}\rvert_{g=1} = \sin^N(\phi/2) + \cos^N(\phi/2) .
\end{align}
Using the Fourier decomposition of the FOTOC, $I^{\hat{S}_x}_m \equiv \sum_{\phi} F_{\phi} e^{-im\phi}$ we obtain
\begin{align}
 I^{\hat{S}_x}_{m}\rvert_{g=1} = \frac{2}{4^N}{2N\choose N-m} , \label{eq:TFI_MQC_exact}
\end{align} 
for even $m$, and is $0$ otherwise. Thus, independent of system size $N$ the critical point of the TFI model is characterized by a Gaussian (binomial) MQC spectrum.

For MQC intensities in the ferromagnetic phase, we find a much simpler analysis proceeds from using the large-$N$ expression for the FOTOC in Eq.~\eqref{eq:TFI_FOTOC_cont}. In fact, this expression yields a constant MQC spectrum for all $g < 1$, and limits to the exact MQC spectrum Eq.~\eqref{eq:TFI_MQC_exact} found at the critical point. From Eq.~\eqref{eq:TFI_FOTOC_cont} we obtain (again, assuming $m$ is even)
\begin{multline}
 I^{\hat{S}_x}_{m}\rvert_{g<1} \simeq \frac{1}{4^N}{2N\choose N-m} \\
 + \frac{2}{\pi 4^N}\sum_{\substack{-N+m\leq j \leq N+m,\\j\,\,\textrm{odd}}}{2N\choose j+N-m}\frac{(-1)^{(j-1)/2}}{j}.
\end{multline}
To write this latter term in a more insightful form, we first apply the approximation
\begin{align}
 {2N \choose j+N-m} \simeq \frac{4^N}{\sqrt{\pi N}} e^{-(j-m)^2/N}.
\end{align}
Since we are generally interested in $m \ll \sqrt{N}$, we may extend the bounds of the sum to $j = \pm\infty$, accumulating an error $\lesssim 2e^{-N}/N$. We then recast our expression as 
\begin{multline}
\frac{2}{\pi 4^N}\sum_{\substack{-N+m\leq j \leq N+m,\\j\,\,\textrm{odd}}}{2N\choose j+N-m}\frac{(-1)^{(j-1)/2}}{j}\\ \simeq \frac{4e^{-m^2/N}}{\pi\sqrt{\pi N}}\sum_{j \mathrm{odd}, j>0} \frac{(-1)^{(j-1)/2}}{j}e^{-j^2/N}.
\end{multline}
Now, the partial sum for all terms with $j \gtrsim \sqrt{N}$ has an absolute value $\lesssim 1/\sqrt{N}$, and neglecting such terms allows us to make the replacement $e^{-j^2/N} \approx 1$. Using $\sum_{j\:\textrm{odd}} (-1)^{(j-1)/2}/j = \pi/4$, and approximating the first term in our MQC intensity with a Gaussian, we arrive at the result
\begin{align}
 I^{\hat{S}_x}_{m}\rvert_{g<1} \simeq \frac{2}{\sqrt{\pi N}}e^{-m^2/N}, \label{eq:TFI_MQC_FM}
\end{align}
which is the large-$N$ equivalent of the exact expression found at the transition.

\Rev{
\subsubsection{Quantitative signatures of QPT}
In the main text we use the derivative $\frac{d^2}{d\Omega^2}I_0^{\hat{S}_x}$ as a quantitative tool to diagnose the QPT of the TFI model. This is motivated by the observation that there is an abrupt change in the MQC spectrum at the critical $(\Omega/\chi)_c$, which is foreshadowed here in our expression for the FOTOC Eq.~\eqref{eq:TFI_FOTOC_cont}. Specifically, plotting the MQC spectrum obtained from Eq.~\eqref{eq:TFI_FOTOC_exact} reveals a kink feature in the MQC intensity $I_0^{\hat{S}_x}$ at the critical point, similar to the abrupt change expected in the order parameter $\langle \vert \hat{S}_z \rangle \vert \rangle$. We quantitatively diagnose this using the associated divergence in the derivative $\frac{d^2}{d\Omega^2}I_0^{\hat{S}_x}$.

In Fig.~\ref{fig:d2scaling}, we plot the onset of this divergence for $N = 100, 300, 1000$ via a discrete Fourier transform of the second derivative of the exact FOTOC expression Eq.~\eqref{eq:TFI_FOTOC_exact}. We observe the development of a delta function-like peak at the critical point as $N\rightarrow\infty$. Despite the general decrease of the FOTOC and MQC intensities for large system sizes, as suggested by scaling of our large $N$-expressions Eq.~\eqref{eq:TFI_FOTOC_cont} and Eq.~\eqref{eq:TFI_MQC_FM}, we nonetheless observe a growth of the second derivative peak with $N$, suggesting a non-vanishing signature of this QPT in the thermodynamic limit.}

\subsection{LMG model}
The LMG model describes a system of $N$ mutually interacting spin-$1/2$s and is defined by the Hamiltonian
\begin{equation}
 \hat{H} = -\frac{\chi}{N}\hat{S}_z^2 - \Omega\hat{S}_x \label{eqn:HLMG}
\end{equation}
where we introduced collective spin operators $\hat{S}_{\alpha} = \sum_j \hat{\sigma}^{\alpha}_j/2$ for $\alpha = x,y,z$. This Hamiltonian emerges from Eq.~(1) of the main text for $\mathcal{C} = N$ and $\chi_{ij} = \chi$.

\subsubsection{Ground state for large $N$}
An exact solution of the ground state and low-energy physics can be obtained via a Holstein-Primakoff transformation in the large $N$ limit. We follow the derivation previously reported in Ref.~\cite{Kwok_2008} and references therein, and summarize the key results here. Specifically, we write the collective spin operators in terms of bosonic operators
\begin{eqnarray}
 \hat{S}_x = \frac{N}{2} - \hat{a}^{\dagger}\hat{a} , \notag \\
 \hat{\tilde{S}}_- \equiv \hat{S}_z - i\hat{S}_y = -\Big(\sqrt{N - \hat{a}^{\dagger}\hat{a}}\Big)\hat{a} .
\end{eqnarray}
Using a $1/N$ expansion of the spin raising and lower operators and restricting to the paramagnetic phase ($\Omega \leq 1$) the LMG Hamiltonian (\ref{eqn:HLMG}) can be written as:
\begin{equation}
 \hat{H}_{\mathrm{HP}} = \frac{1}{2}(2\Omega-\chi)\hat{a}^{\dagger}\hat{a} -\frac{\chi}{4}\left( \hat{a}^2 + \hat{a}^{\dagger 2}\right) .
\end{equation}
This bosonic Hamiltonian is then diagonalized via the Bogoliubov transformation $\hat{a} = \mathrm{sinh}(r)\hat{b}^{\dagger} + \mathrm{cosh}(r)\hat{b}$ with $\mathrm{tanh}(2r) = \chi/(2\Omega-\chi)$, such that $\hat{H}_{\mathrm{HP}}$ becomes:
\begin{equation}
 \hat{H}_{\mathrm{bog}} = \sqrt{\Omega(\Omega-\chi)}\hat{b}^{\dagger}\hat{b}.
\end{equation}
The eigenstates of $\hat{H}_{\mathrm{bog}}$ are Fock states of the Bogoliubons, $\hat{b}^{\dagger}\hat{b}\vert n \rangle = n \vert n \rangle$. As the ground state is then the vacuum of the Bogoliubons, $\vert \psi_{\mathrm{gs}} \rangle_{\mathrm{bog}} = \vert 0\rangle$, then consequently the ground state of the paramagnetic phase in the original Holstein-Primakoff bosons is the squeezed vacuum \cite{walls_quantum_2008},
\begin{equation}
 \vert \psi_{\mathrm{GS}} \rangle_{\mathrm{HP}} = \frac{1}{\sqrt{\mathrm{cosh}(r)}}\sum_{n=0}^{\infty} \mathrm{tanh}(r)^n \frac{\sqrt{(2n)!}}{2^n n!} \vert 2n \rangle . \label{eqn:gs_HP}
\end{equation}

\subsubsection{MQC spectrum}
The MQC spectrum of the ground state in the paramagnetic phase of the LMG model can be computed exactly using the result of $\vert \psi_{\mathrm{GS}} \rangle_{\mathrm{HP}}$. Specifically, the spectrum with respect to the basis of $\hat{S}_x$ can be computed via the mapping $\hat{S}_x = \frac{N}{2} - \hat{a}^{\dagger}\hat{a}$, and using the definition of the MQC intensities in terms of the Fourier series
\begin{equation}
 F_{\phi} \equiv \vert \langle \psi_{\mathrm{GS}} \vert e^{-i\phi\hat{S}_x} \vert\psi_{\mathrm{GS}} \rangle \vert^2 = \sum_m I^{\hat{S}_x}_m(\hat{\rho}_{\mathrm{GS}}) e^{-im\phi} .
\end{equation}
The LHS is solved upon substitution of Eq.~(\ref{eqn:gs_HP}) and we can then identify by inspection:
\begin{multline}
 I^{\hat{S}_x}_m(\hat{\rho}_{\mathrm{GS}}) = \frac{m!}{2^m [(m/2)!]^2} \\
 \times {}_2F^1\left( \frac{1}{2}, \frac{1+m}{2}; \frac{2+m}{2}; \mathrm{tanh}(r)^4\right) \mathrm{tanh}(r)^m , \label{eqn:ImLMGexact}
\end{multline}
where ${}_2F^1(a,b;c;x)$ is a hypergeometric function. Equation (\ref{eqn:ImLMGexact}) is plotted in Fig.~2 of the main text for $m=0,2$. 

As they approach the critical point $(\Omega/\chi)_c = 1$ from the paramagnetic phase the MQC intensities are found to vanish. Specifically, Eq.~(\ref{eqn:ImLMGexact}) can be expanded in the vicinity of $(\Omega/\chi)_c = 1^+$ as
\begin{multline}
 \left. I^{\hat{S}_x}_m(\hat{\rho}_{\mathrm{GS}})\right\vert_{\frac{\Omega}{/\chi} \to 1^+} \approx \frac{2}{\pi} \sqrt{\frac{\Omega}{\chi} - 1} \left[ 2H_{\frac{m-1}{2}} 
 - 2\mathrm{log}(2) \right. \\
 \left. - \mathrm{log}\left(\frac{\Omega}{\chi} - 1\right) \right] . \label{eqn:ImVanish_HP}
\end{multline}
where $H_n = \sum_{k=1}^n (1/k)$ is the $n$th Harmonic number. We stress that even though $I^{\hat{S}_x}_m(\hat{\rho}_{\mathrm{GS}}) \to 0$ as $(\Omega/\chi)_c \to 1^+$, the analytical solution explicitly preserves $\sum_m I^{\hat{S}_x}_m(\hat{\rho}_{\mathrm{GS}})=1$ for $N \to \infty$ (as we have assumed in the original derivation of the ground state).

The vanishing of the MQC spectrum as $N\to\infty$ is reconciled by the fact that we expect the spectrum to become infinitely broad. Although our $N\to\infty$ calculation does not facilitate calculation of the MQC spectrum in the ferromagnetic phase, we can gain insight by computing it exactly for the ferromagnetic ground state $\vert \psi^{F}_{\mathrm{GS}} \rangle = (\vert (N/2)_z \rangle + \vert -(N/2)_z \rangle)/\sqrt{2}$ for arbitrary $N$. Specifically, using the identity
\begin{multline}
 \langle (-N/2 + q)_z \vert e^{-i\phi\hat{S}_x} \vert (-N/2)_z \rangle = \\
 (-1)^q \sqrt{\frac{N!}{(N-q)!q!}} \mathrm{cos}^{N-q} (\phi/2) \mathrm{sin}^q(\phi/2) 
\end{multline}
we obtain 
\begin{multline}
 \vert \langle \psi^{F}_{\mathrm{GS}} \vert e^{-i\phi\hat{S}_x} \vert\psi^{F}_{\mathrm{GS}} \rangle \vert^2 = \\ \mathrm{cos}^{2N}(\phi/2) 
 + \mathrm{sin}^{2N}(\phi/2) + 2^{1-N}\mathrm{sin}^N(\phi) ,
\end{multline}
and 
\begin{equation}
 I^{\hat{S}_x}_m(\hat{\rho}^F_{\mathrm{GS}}) = \frac{2}{4^N} \binom{2N}{N-m} + \frac{2(-1)^{\frac{m}{2}}}{4^N} \binom{N}{\frac{N-m}{2}}
\end{equation}
for $m$ even and $I^{\hat{S}_x}_m(\hat{\rho}^F_{\mathrm{GS}}) = 0$ otherwise. From this last result we similarly observe the individual intensities of the ferromagnetic ground state should vanish as $N\to\infty$. 

\begin{figure*}[tb]
 \includegraphics[width=16cm]{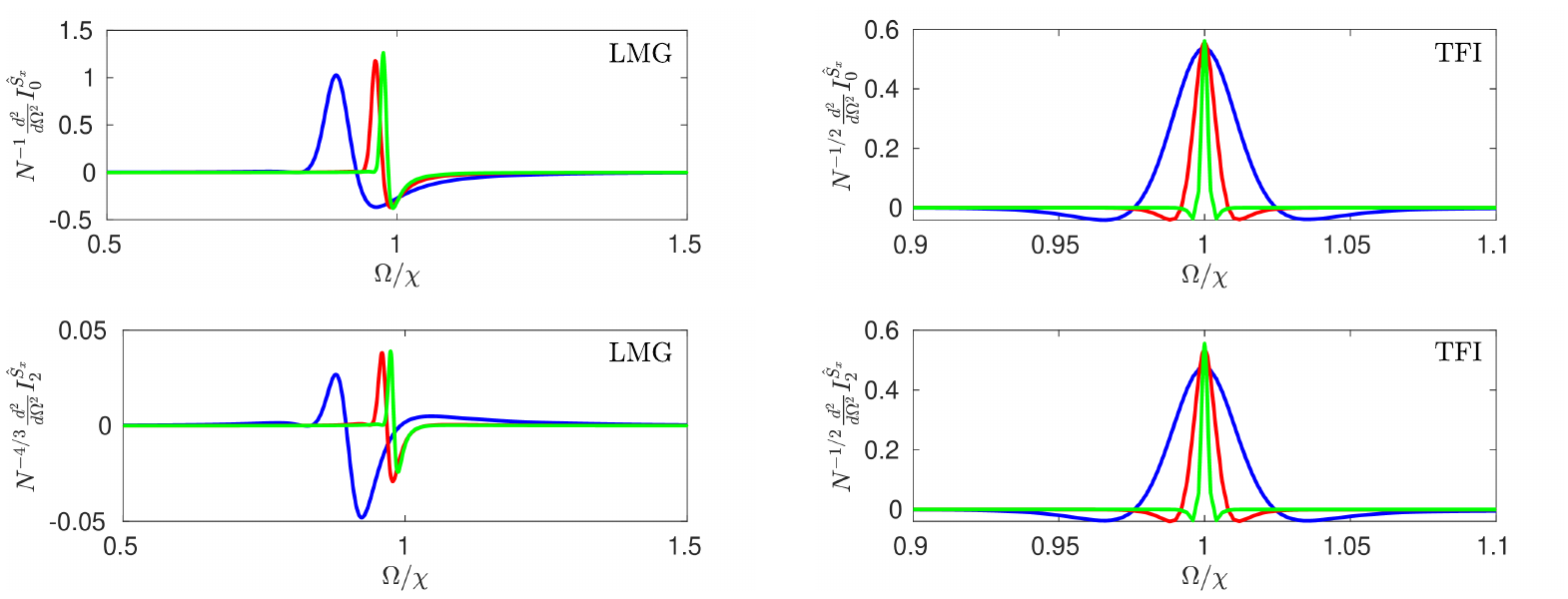}
 \caption{\Rev{Emergence of QPT signature in the derivative $\frac{d^2}{d\Omega^2} I^{\hat{S}_x}_{0,2}$ for LMG (left) and TFI (right) models. Data is for: $N=100$ (blue), $N=500$ (red) and $N=1000$ (green) for the LMG model, and $N=100$ (blue), $N=300$ (red) and $N=1000$ (green) for the TFI model. We normalize the derivatives with respect to the approximate scaling of the divergence approaching from the ferromagnetic phase [i.e., $(\Omega/\chi)_*$ corresponds to the positive peak in the LMG model], $\frac{d^2}{d\Omega^2}I^{\hat{S}_x}_0\vert_{(\Omega/\chi)_*} \propto N$ and $\frac{d^2}{d\Omega^2}I^{\hat{S}_x}_2\vert_{(\Omega/\chi)_*} \propto N^{4/3}$ for the LMG model, and $\frac{d^2}{d\Omega^2}I^{\hat{S}_x}_{0,2}\vert_{(\Omega/\chi)_*} \propto N^{1/2}$ for the TFI model.}}
 \label{fig:d2scaling}
\end{figure*}

\Rev{
From this collection of results we thus propose that the QPT is signaled in the MQC spectrum by a singularity of the second derivative of the individual intensities, $\frac{d^2}{d\Omega^2} I^{\hat{S}_x}_m$, at the critical point $(\Omega/\chi) \to 1$. A practical quantitative signature of this is the divergence of $\frac{d^2}{d\Omega^2} I^{\hat{S}_x}_m$ as $(\Omega/\chi) \to 1^{\pm}$. For LMG model we are able to analytically compute this former limit using the $N\to\infty$ calculation [building on Eq.~(\ref{eqn:ImVanish_HP}], which predicts a divergence to minus infinity when approaching the QPT from the paramagnetic phase as:
\begin{multline}
 \left.\frac{d^2I^{\hat{S}_x}_m(\hat{\rho}_{\mathrm{GS}})}{d\Omega^2}\right\vert_{\frac{\Omega}{\chi} \to 1^+} \approx \frac{1}{2\pi\left(\frac{\Omega}{\chi} - 1\right)^{3/2}}\left[ \mathrm{log}(4) - 2H_{\frac{m-1}{2}} \right. \\ \left.
 + \mathrm{log}\left(\frac{\Omega}{\chi} - 1 \right) \right].
\end{multline}
In Fig.~\ref{fig:d2scaling} we plot the development of the discontinuity for finite systems $N=100,500,1000$ by numerically computing the MQC spectrum, as per the main text. We observe that a divergence to $\pm \infty$ is developing as $N\to\infty$ depending on how the critical point is approached. 
}

\Rev{
\section{Finite size effects}
To further verify that the MQC intensities bear signatures of the QPT we must account for systematic offsets and defects due to finite system size. Of particular importance is the deviation of the identified phase boundary for a finite system relative to the actual QPT critical point obtained in the thermodynamic limit. Here, we present a brief investigation of finite size effects with regards to the MQC intensities of the LMG and TFI models.

\begin{figure}[tb]
 \includegraphics[width=8cm]{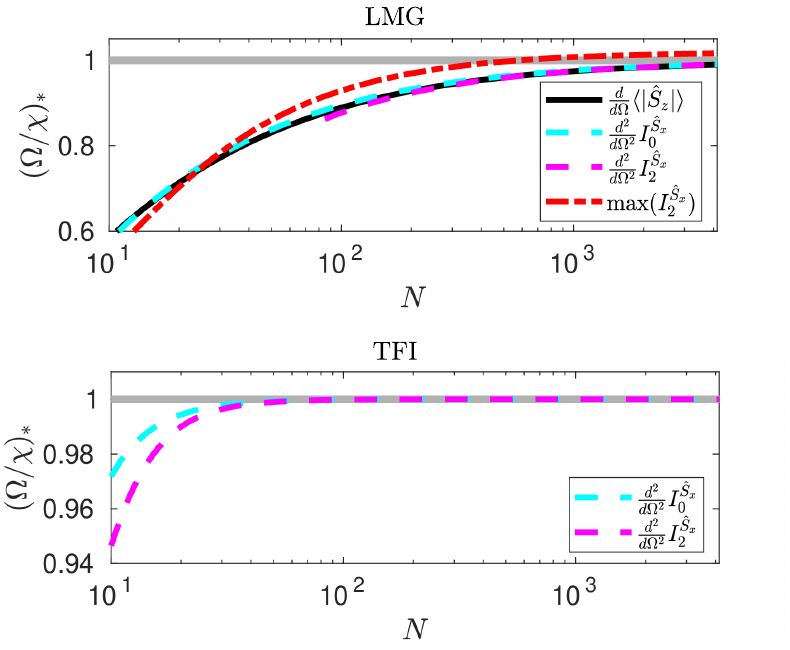}
 \caption{\Rev{Finite size effects in the critical point $(\Omega/\chi)_*$ for LMG (top panel) and TFI (bottom panel) models. A fit to to the location of the positive peak in $\frac{d^2}{d\Omega^2}I^{\hat{S}_x}_0$ indicates that it approaches the QPT at $\Omega_c = 1$ as $1-(\Omega/\chi)_* \sim N^{-0.65}$ and $1-(\Omega/\chi)_* \sim N^{-2}$ for the LMG and TFI models respectively. For comparison, a fit to the location of the peak in the susceptibility $d\langle\vert\hat{S}_z\vert\rangle$ gives $1-(\Omega/\chi)_* \sim N^{-0.65}$ for the LMG model. Fits are obtained from data across $200 < N < 5000$.}}
 \label{fig:ImScaling}
\end{figure}

Figure 4 presents the approach of the phase boundary $(\Omega/\chi)_*$, determined by a number of methods which we discuss momentarily, with respect to the QPT at $(\Omega/\chi)_c = 1$. We identify the location of the phase boundary using four independent signatures: i) the peak of the susceptibility $\frac{d}{d\Omega} \langle \vert \hat{S}_z \vert \rangle$, ii) the peak of the second derivative $\frac{d^2}{d^2\Omega} I^{\hat{S}_x}_{0}$, iii) the peak of $\frac{d^2}{d^2\Omega} I^{\hat{S}_x}_{2}$ and iv) the maximum of the cusp which appears in the $m=2$ intensity as a function of $\Omega$, i.e., $\mathrm{max}(I^{\hat{S}}_2)$ (see Fig.~2 and the main text for discussion of this feature). For ii) and iii) we clarify that the peak is taken to be the positive valued feature in $\frac{d^2}{d^2\Omega} I^{\hat{S}_x}_{0,2}$ as discussed in the previous section.

As shown in Fig. \ref{fig:ImScaling}, for the LMG model i)-iii) approach the true QPT faithfully as $N$ is increased with similar scaling of $1 - (\Omega/\chi)_* \sim N^{-0.65}$, $1 - (\Omega/\chi)_* \sim N^{-0.65}$ and $1 - (\Omega/\chi)_* \sim N^{-0.68}$ respectively. The scaling exponents of the magnetic susceptibility and $\frac{d^2}{d^2\Omega} I^{\hat{S}_x}_{0}$ agree within the error margins of our fit over the window $200 < N < 5000$. The cusp $\mathrm{max}(I^{\hat{S}}_2)$ is similar qualitatively, although for large $N$ it clearly converges to a point adjacent to the QPT in the paramagnetic phase, $(\Omega/\chi)_* > (\Omega/\chi)_c$. Nevertheless, we believe this feature may still be useful to approximately locate the phase boundary for experiments which do not have the technical capability to resolve the features of $\frac{d^2}{d^2\Omega} I^{\hat{S}_x}_{0}$. Lastly, we note that the peak value of the derivatives scale approximately as $\frac{d^2}{d\Omega^2}I^{\hat{S}_x}_0\vert_{(\Omega/\chi)^+_*} \propto N$ and $\frac{d^2}{d\Omega^2}I^{\hat{S}_x}_2\vert_{(\Omega/\chi)^+_*} \propto N^{4/3}$.

For the TFI model, analytical and numerical limitations preclude the evaluation of i) for meaningful system sizes, and iv) is irrelevant, as this cusp is found to be uncorrelated with the transition and thus, unlike the LMG model, cannot be used as a proxy for the phase boundary in lieu of ii-iii). However, as shown in Fig. \ref{fig:ImScaling} we still find that both ii) and iii) similarly approach the critical point as $1 - (\Omega/\chi)_* \sim N^{-2}$, within error margins over $200 < N < 5000$. The deviation of these signatures from the true critical point on the ferromagnetic side of the transition is largely consistent with finite size deviations found in exact studies of the peak of the fidelity susceptibility for this model \cite{Damski_2013}, which also approaches the transition from the ferromagnetic side as $1-(\Omega/\chi)_* \sim N^{-2}$. We further observe that the magnitude of both ii) and iii) scale as $\sim N^{1/2}$
over the same window, with a full width at half maximum scaling as $~1/N$, implying a general improvement in signal resolution of these signatures for larger system sizes, as can be observed by the sharpening peak in Fig.~\ref{fig:d2scaling}.
}

\Rev{
\section{Relevance of decoherence}
Decoherence is an important practical factor to consider when obtaining FOTOCs from quasi-adiabatic ramps. At first-order, decoherence impacts FOTOCs and the MQC spectrum by limiting the total time over which the Hamiltonian parameters can be slowly quenched and thus the degree to which diabatic excitations can be minimized. However, decoherence also directly degrades the MQC intensities, as has been shown in recent experimental and theoretical work focusing on quench dynamics in trapped-ion quantum simulators \cite{Martin2017_OTOC,Martin_2018}. In the main text we effectively focused on the former by assessing how diabatic excitations impact the quality of the FOTOCs and MQC spectrum we obtain. Here, we briefly address the relevance of decoherence in state-of-the-art trapped-ion experiments, which could be used to simulate the LMG and TFI models. In particular, we contrast the timescales required for effective quasi-adiabatic ramps against typical timescales for decoherence. 

For the LMG model we can estimate the relevance of decoherence by comparing to the experiments reported in Refs.~\cite{Bollinger_2018_arxiv,Bollinger_2018}. These experiments, based on a $2$D Penning trap trapped-ion simulator, studied quasi-adiabatic ramps in the closely related Dicke model which describes the interaction of a single collective spin and bosonic mode. The LMG model we study emerges in the limit where the bosons can be adiabatically eliminated, i.e. when the boson dynamics are fast relative to the spin dynamics. In that case, the bosons essentially act passively to mediate effective all-to-all spin-spin interactions. Using the proposed parameters from Ref.~\cite{Bollinger_2018_arxiv} we estimate the spin-spin interaction $\chi$ that could be reasonably achieved: $\chi = g^2/\delta \approx 2\pi \times 810$~Hz where $g = 2\pi \times 1.32$~kHz is the spin-boson coupling and $\delta = 2\pi \times 4$~kHz is the detuning of the boson mode. 

Comparatively, it is quoted in Ref.~\cite{Bollinger_2018_arxiv} that with modest near-term technical improvements it is possible to limit single-particle spin decoherence rates, $\Gamma$, to as low as $\Gamma \approx 15$s$^{-1}$. Combining this estimate with the previous value for $\chi$ ($\Omega$, implemented by microwaves, is effectively completely tunable over the relevant parameter regime) Ref.~\cite{Bollinger_2018_arxiv} projects that suitably designed quasi-adiabatic ramps with a typical timescale of $1$-$4$ms could achieve ground state fidelities on the order of $0.4$-$0.8$ even accounting for decoherence and technical noise. Such fidelities would be more than sufficient to observe the features of the MQC spectrum we report in this manuscript. Moreover, while the system size discussed in Ref.~\cite{Bollinger_2018_arxiv} is a little smaller than we include in this manuscript ($N=50$ in Fig.~3), we expect any possible decrease in the projected fidelity with increased system size can be strongly offset by recent improvements in increasing the spin-boson interaction $g$ (and thus spin-spin interaction $\chi$) \cite{Ge_2019}.

We expect similar results can be achieved in the near-term for other spin models with shorter-ranged interactions, which may also be simulated in trapped-ion platforms. For example, decoherence in $1$D ion chains, which have been used to simulate spin models with power-law interactions \cite{Kim2009,Jurcevic2014}, is typically far lower relative to the $2$D ion crystal discussed above.

Lastly, we comment on the role of decoherence in our choice to focus on low-$m$ intensities, such as $I^{\hat{S}_x}_0$ and $I^{\hat{S}_x}_2$, in the manuscript. Foremost, these intensities will be the easiest to measure experimentally as they feature a large signal-to-noise ratio. In comparison, intensities for larger $m$ have a reduced signal-to-noise ratio and are thus susceptible to errors due to technical noise and decoherence. In particular, obtaining reliable estimates of the larger-$m$ intensities from quasi-adiabatic echoes, relative to the expectation for the true ground state, will place more stringent requirements on the adiabaticity of the protocol than those required for obtaining low-$m$ intensities. These longer ramps, in turn, exacerbate the effects of decoherence and reduce the available experimental signal (roughly, we expect the intensities to be reduced by a factor $\propto e^{-N\Gamma \tau}$ \cite{Martin_2018}). Therefore, this makes them much harder to access experimentally \cite{Martin2017_OTOC,Martin_2018}. This limitation can become relevant if one wants to focus on the use of the MQC spectrum to characterize the QFI and with it the QPT, whereas the versatility of our scheme allows us to also use the low-$m$ intensities as signatures of the QPT. 
}

\Rev{
\section{Examples for Non-Integrable Models}
In the main text, we demonstrated the use of FOTOCs and the associated MQC intensities for detecting QPTs in two of the conceptually simplest, paradigmatic models exhibiting QPTs: the LMG model and the TFI model. Both of these models are exactly solvable, and are similarly characterized by Ising interactions with the addition of a transverse field. This facilitated a detailed analysis and comparison against well-known results. Here, we extend our scope to briefly consider a few examples of more complex, non-integrable models and study the MQC spectrum of their corresponding ground states and the relation to potential QPTs.

\subsection{Axial-next-nearest-neighbor Ising model}
We start with the axial next-nearest-neighbor Ising (ANNNI) model \cite{Chakrabarti1996}, a simple extension of the TFI model which includes next-nearest-neighbor (nnn) coupling terms,
\begin{equation}
 \hat{H}_{\rm{ANNNI}} = -\chi\sum_{i=1}^N\hat{\sigma}^z_i\hat{\sigma}^z_{i+1} -\gamma\sum_{i=1}^N\hat{\sigma}^z_i\hat{\sigma}^z_{i+2} - \Omega\sum_{i=1}^N\hat{\sigma}^x_i . \label{eqn:Hanni}
\end{equation}
Here, $\gamma$ characterizes the nnn coupling strength: $\gamma = 0$ reduces (\ref{eqn:Hanni}) to the usual TFI model. Qualitatively, the effect of the nnn terms is to either further reinforce the ferromagnetic order from the Ising interactions when $\gamma/\chi > 0$, pushing the phase boundary to larger transverse field strengths, or to destabilize this order when $\gamma/\chi < 0$. The ANNNI model is relevant for modelling the effects of imperfect spin-spin interactions in a trapped-ion implementation of the TFI model. In such a setup, the strength of the nnn interaction is correlated to the nearest-neighbor interaction strength, $\chi$, and remains independent of the transverse field strength, $\Omega$. We thus consider a parameter $\lambda = \Omega/\chi$ similar to the main text and tune $\Omega$, leaving $\chi$ and $\gamma$ fixed.

\begin{figure}[tb!]
 \includegraphics[width=8cm]{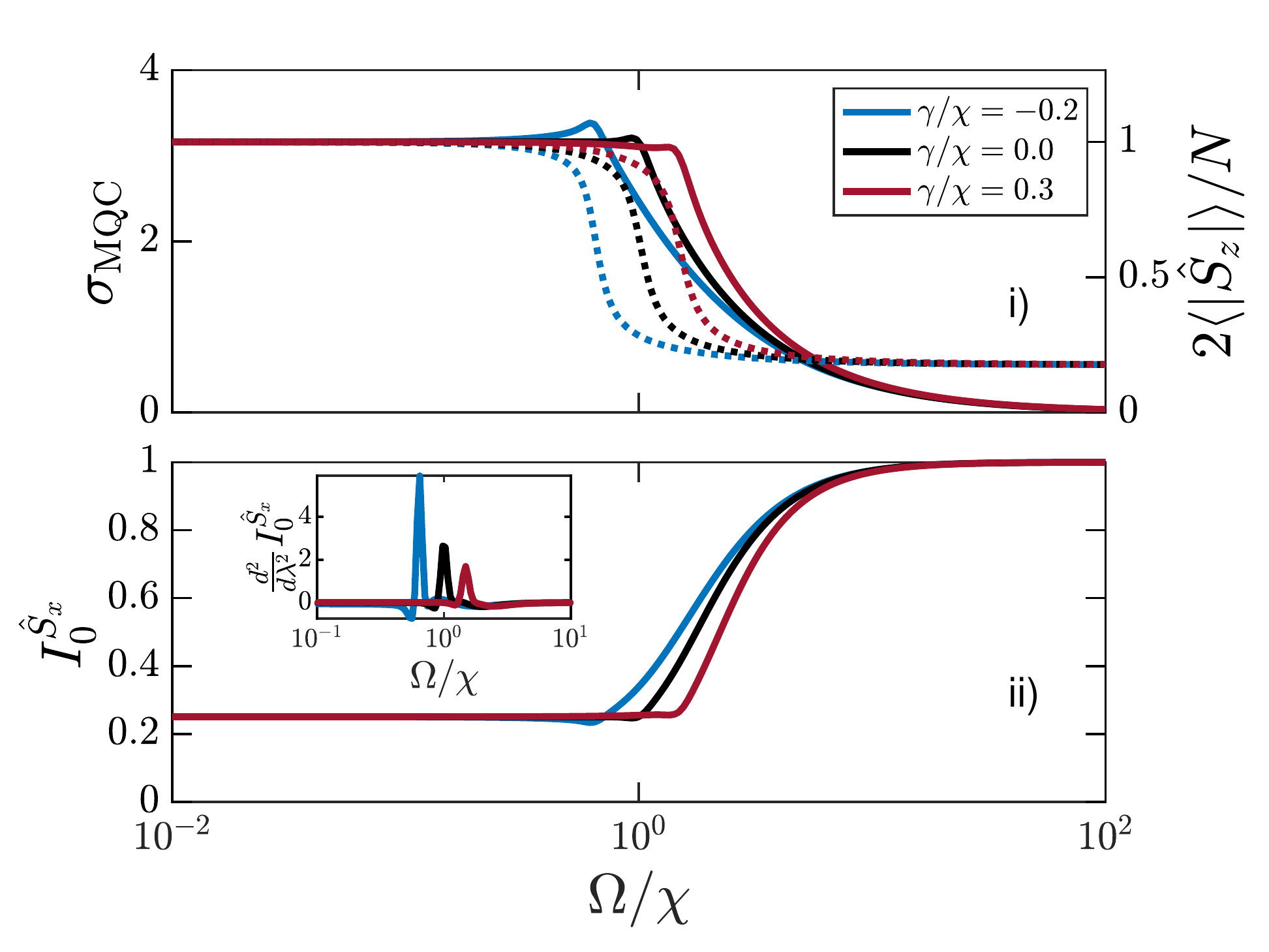}
 \includegraphics[width=8cm]{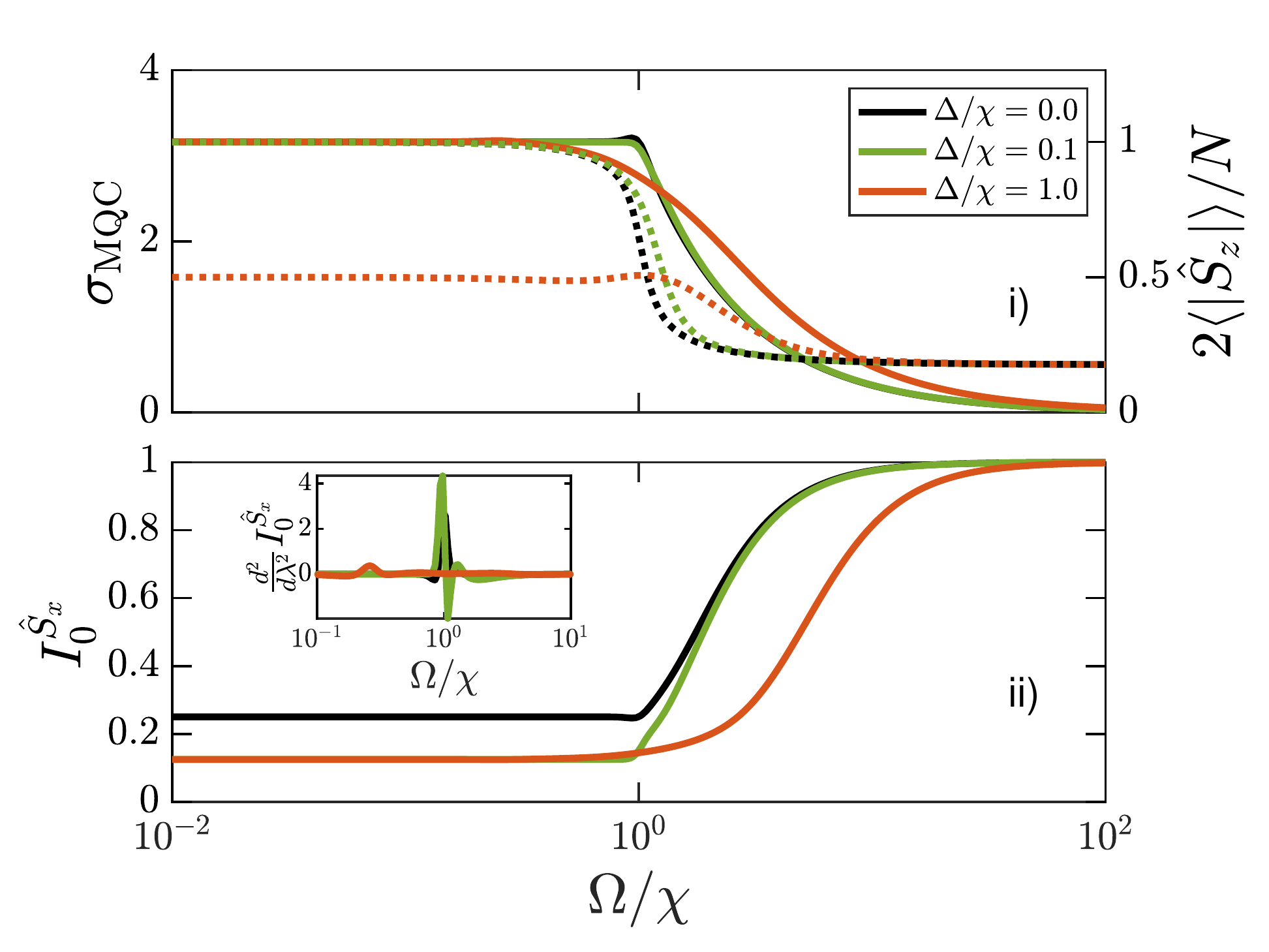}
 \caption{\Rev{Aspects of the characteristic MQC spectrum $I^{\hat{S}_x}_m(\hat{\rho})_{\rm{G.S.}}$ for the ANNNI model (top panel), and for the RFTI model (bottom panel) averaged over 100 random disorder realizations. All ground states numerically-computed for $N=20$ via a Lanczos algorithm. i) The MQC spectrum width, $\sigma_{\rm{MQC}}$ (solid lines) and order parameter, $2\langle|\hat{S}_z|\rangle/N$, (dotted lines) for select values of the nnn coupling strength $\gamma/\chi$ for the ANNNI model, and of the random disorder strength $\Delta/\chi$ for the RFTI model. ii) The corresponding $I_0^{\hat{S}_x}$ intensities, ii) (inset) as well as the associated second derivatives with respect to the parameter $\lambda = \Omega/\chi$.}}
 \label{fig:ANNNIa}
 \label{fig:RFIb}
\end{figure}

We numerically compute the exact ground state of the ANNNI Hamiltonian for $N=20$, and plot the resulting MQC spectrum $I^{\hat{S}_x}_m$ in Fig.~\ref{fig:ANNNIa}, again with respect to the ideal $\lambda \rightarrow \infty$ Hamiltonian, $\hat{S}_x$. Similar to our analysis of the TFI model (shown here as $\gamma/\chi = 0$), we plot the width of the MQC spectrum $\sigma_{\rm{MQC}} = (\sum_m m^2I^{\hat{S}_x}_m)^{1/2}$ and the order parameter $2\langle |\hat{S}_z|\rangle/N$ in Fig.~\ref{fig:ANNNIa}. We also plot the central MQC component $I_0^{\hat{S}_x}$ in Fig.~\ref{fig:ANNNIa} for various values of $\gamma/\chi$. Utilizing the peak of the second derivative of $I_0^{\hat{S}_x}$ with respect to $\lambda$ to locate the ``kink'' in this component, which we use to locate the phase boundary, we find critical points $(\Omega/\chi)_c = 0.64$, $0.98$, and $1.48$ for $\gamma/\chi = -0.2$, $0.0$, and $0.3$, respectively. These are reasonably close to the expected transition locations of $(\Omega/\chi)_c \approx 0.65$, $1.00$, and $1.38$ \cite{Chakrabarti1996,Karrasch2013}, though we note increasing deviations for larger values of $\gamma/\chi$. However, we also note that for larger $\gamma/\chi$ the signal in the second derivative becomes less resolvable, indicating larger sensitivity to finite size effects for these models, which may account for the respective deviations from the expected critical point. However, a full investigation of the scaling of the MQC spectra of such models is beyond the scope of this work.

\subsection{Random-field transverse Ising model}
We similarly compute the ground state of the random field transverse Ising (RFTI) model,
\begin{equation}
\hat{H}_{\rm{RFTI}} = -\chi\sum_{i=1}^N\hat{\sigma}^z_i\hat{\sigma}^z_{i+1} -\sum_{i=1}^N\delta_i\hat{\sigma}^z_i - \Omega\sum_{i=1}^N\hat{\sigma}^x_i ,
 \label{eqn:HRFI}
\end{equation}
where $\delta_i$ is a Gaussian random variable with zero mean and variance $\Delta^2$. We again compute the MQC spectrum with respect to the operator $\hat{S}_x$, under the assumption that the random disorder in the model is not inherently linked to the strength of the transverse field, i.e. $\Omega/\chi$ can be tuned independently of $\Delta/\chi$.

To numerically compute the MQC spectrum and obtain reasonably smooth results for the corresponding derivatives, we use fixed disorder realizations for all values of $\Omega/\chi$, and average the resulting spectra over several such realizations (see Fig.~\ref{fig:RFIb}). In general, the RFTI model does not order in $1$D \cite{Chakrabarti1996}. However, in the presence of sufficiently weak disorder the finite extent of the system in our simulations ($N=20$) remains the most relevant factor, and signals of the ferromagnetic to paramagnetic QPT in the clean system should remain. Indeed, for $\Delta/\chi = 0.1$, the sharp peak in the second derivative remains at the location of the clean system transition. However, for larger disorder this peak vanishes, as do the visible kinks in the $I_0^{\hat{S}_x}$ component and spectrum width $\sigma_{\rm{MQC}}$. The signatures identified in the MQC spectra are thus reliable indicators of the presence or absence of underlying critical phenomena for a much broader swath of spin models and possible terms, and can be used as indicators for QPTs regardless of in-depth knowledge of the underlying terms in the Hamiltonian or prior knowledge of the nature of the transition and related order parameters.
}